\documentclass{emulateapj}
\usepackage{color} 

\usepackage{epsfig}
\usepackage{subfigure}
\usepackage{graphicx}
\usepackage{amsmath}
\usepackage{natbib}
\usepackage{lscape}
\newcommand{\pa}{\partial}
\newcommand{\mb}{\boldsymbol}
\newcommand{\mc}{\mathcal}

\shorttitle{Wind-driven Accretion in PPDs}
\shortauthors{X.-N. Bai \& J. M. Stone}

\begin{document}


\title{Wind-driven Accretion in Protoplanetary Disks --- I: Suppression
of the Magnetorotational Instability and Launching of the Magnetocentrifugal Wind}


\author{Xue-Ning Bai\altaffilmark{1,2,3} \& James M. Stone\altaffilmark{1}}

\altaffiltext{1}{Department of Astrophysical Sciences, Peyton Hall, Princeton
University, Princeton, NJ 08544}
\altaffiltext{2}{Institute for Theory and Computation,
Harvard-Smithsonian Center for Astrophysics, 60 Garden St., MS-51, Cambridge, MA 02138}
\altaffiltext{3}{Hubble Fellow}
\email{xbai@cfa.harvard.edu}




\begin{abstract}
We perform local, vertically stratified shearing-box MHD simulations of protoplanetary disks
(PPDs) at a fiducial radius of 1 AU that take into account the effects of both Ohmic resistivity
and ambipolar diffusion (AD). The magnetic diffusion coefficients are evaluated
self-consistently from a look-up table based on equilibrium chemistry. We first show that the
inclusion of AD dramatically changes the conventional picture of layered accretion. Without
net vertical magnetic field, the system evolves into a toroidal field dominated configuration
with extremely weak turbulence in the far-UV ionization layer that is far too inefficient to drive
rapid accretion. In the presence of a weak net vertical field (plasma $\beta\sim10^5$ at
midplane), we find that the magnetorotational instability (MRI) is completely suppressed,
resulting in a fully laminar flow throughout the vertical extent of the disk. A strong
magnetocentrifugal wind is launched that efficiently carries away disk angular momentum and
easily accounts for the observed accretion rate in PPDs. Moreover, under a physical disk wind
geometry, all the accretion flow proceeds through a strong current layer with thickness of
$\sim0.3H$ that is offset from disk midplane with radial velocity of up to $0.4$ times the sound
speed. Both Ohmic resistivity and AD are essential for the suppression of the MRI and wind
launching. The efficiency of wind transport increases with increasing net vertical magnetic flux
and the penetration depth of the FUV ionization. Our laminar wind solution has important
implications on planet formation and global evolution of PPDs.
\end{abstract}



\section{Introduction}\label{sec:intro}

Protoplanetary disks (PPDs) are gaseous disks surrounding protostars. The
gas in PPDs are found to be rapidly accreting to the protostar with accretion
rate of $10^{-8\pm1}M_{\bigodot}$ yr$^{-1}$, with typical disk lifetime of about
1-10 Myrs (e.g., \citealp{Hartmann_etal98,Sicilia_etal06}). Despite large number
of observational programs aiming at revealing the structure, composition and
evolution of PPDs (see \citealp{WilliamsCieza11} and references therein), two
crucial theoretical questions on the gas dynamics of PPDs remain poorly
understood: What is the level of turbulence in PPDs? How efficient is angular
momentum transport in PPDs? The answer to these questions are essential to
understanding the structure and evolution of the PPDs, as well as a series of
processes in planet formation. In particular, grain growth (e.g.,
\citealp{Birnstiel_etal10,Zsom_etal10}), transport of solids (e.g.,
\citealp{Garaud07,HughesArmitage10}) are both sensitive to the radial structure
of PPDs and level of turbulence. Current models for planetesimal formation such
as the streaming instability
\citep{Johansen_etal09,BaiStone10b,BaiStone10c} and gravitational instability
\citep{Youdin11} generally favor weak turbulence and small radial pressure
gradient. Moreover, turbulent mixing and disk winds have a significant influence
on the disk chemistry \citep{Semenov_etal06,Heizeller_etal11}. When planets
have formed, planet migration via planet-disk interaction also depends
on the disk radial profile and diffusion processes
(e.g., \citealp{Paardekooper_etal10,Baruteau_etal11}).

\subsection[]{The Current Understanding of Accretion in PPDs}

We are mainly interested in the T-Tauri (class II) phase of PPDs when the
envelope infall has ended and the entire disk is visible. At this stage the
disk is in general not massive enough for gravitational instability to take
place \citep{Zhu_etal10b}. In this paper, we focus on magnetic mechanisms.

Most regions of PPDs\footnote{The region that is close to the inner edge of
PPDs is sufficiently hot ($\gtrsim10^3$K) due to direct illumination by the star
that thermal ionization of Alkali species Na and K will provide sufficient
ionization and the gas behave as ideal MHD \citep{UmebayashiNakano88},
which is not the concern of this paper.}
are very weakly ionized \citep{Hayashi81,IG99}, hence the gas dynamics is
strongly affected by non-ideal magnetohydrodynamics (MHD) effects due to
the finite gas conductivity, which include Ohmic resistivity, Hall effect and
ambipolar diffusion (AD). All three effects are relevant and
important in PPDs \citep{Wardle07,Bai11a}. Generally speaking, Ohmic
resistivity dominates in dense regions with weak magnetic field (e.g., midplane
in the inner region of PPDs), AD dominates in tenuous regions with strong
magnetic field (e.g., disk surface and outer region of PPDs), while the Hall
regime lies in between.

It is widely believed that PPDs are turbulent as a result of the magnetorotational
instability (MRI, \citealp{BH98}). The MRI turbulence transports angular momentum
radially within the disk that allows the majority of the materials to be accreted onto the
protostar while a small fraction of mass disperses away. Non-ideal MHD effects in
PPDs strongly modify the behavior of the MRI. Currently, most studies of the MRI in
PPDs take into account only the effect of Ohmic resistivity, and it is found that in the
inner region of PPDs (about 0.5-5 AU), the disk midplane is too weakly ionized for
the MRI to operate (i.e., the dead zone, \citealp{Gammie96}), while the disk surface is
still prone to the MRI and should be turbulent (i.e., the active layer). A large number of
numerical simulations have been conducted either in the local shearing-box framework
or using full global approach to study and characterize the gas dynamics of the active
layer and the dead zone, as well as exploring their physical consequences (e.g.,
\citealp{FlemingStone03,Turner_etal07,TurnerSano08,IlgnerNelson08,OishiMacLow09,
Dzyurkevich_etal10,HiroseTurner11,Flaig_etal12,Flock_etal12}).
These studies show that the boundary between the active layer and the dead zone is
characterized by the Ohmic Elsasser number $\Lambda\approx1$ (see equation
(\ref{eq:elsasser}) ). Moreover, the dead zone is not completely ``dead" in the sense
that sound waves injected from the active layer bounce back and forth and give rise
to some small Reynolds stress. Nevertheless, angular momentum transport is largely
dominated by the MRI turbulence in the active layer, and the strength of the
turbulence (considering Ohmic resistivity only) appears to be able to drive rapid
accretion consistent with observations.

Comparatively, the effects of non-ideal MHD terms other than Ohmic resistivity on
the MRI in PPDs are less well understood. They are extensively studied in the linear
regime without vertical stratification (see
\citealp{Wardle99,BalbusTerquem01,WardleSalmeron12} for the Hall effect, 
\citealp{BlaesBalbus94,KunzBalbus04,Desch04} for the effect of AD, and
\citealp{PandeyWardle12} for a general study), and with vertical stratification
\citep{SalmeronWardle05,SalmeronWardle08}. In the non-linear regime, so far all
numerical simulations (all using the local shearing-box approach) focus on
individual non-ideal MHD effects such as MRI with Ohmic and Hall terms
\citep{SanoStone02a,SanoStone02b} and MRI with AD
\citep{HawleyStone98,BaiStone11}. Most of the simulations are vertically unstratified.
These simulations provide useful criteria for the MRI to be self-sustained in the
non-linear regime and the results were applied in the framework developed by
\citet{Bai11a} to estimate the efficiency of MRI-driven angular momentum transport in
PPDs. It was shown that even in the most optimal scenario, the MRI-driven accretion
rate falls below typical observed rate by about an order of magnitude at the inner disk
around 1 AU. The main reason is that the strength of the MRI turbulence is expected
to be dramatically reduced in the conventional ``active layer" of the disk once AD is
taken into account. Similar conclusions were also drawn from
\citet{PerezBeckerChiang11a,PerezBeckerChiang11b}, and from
\citet{Mohanty_etal13,Dzyurkevich_etal13} for different stellar masses. Relatively
large accretion rate can be achieved in the outer region of the disk under optimal
magnetic field geometry, and with the assistance of tiny grains \citep{Bai11b}. 


An alternative scenario for describing the gas dynamics in PPDs is the picture of
magnetocentrifugal wind \citep{BlandfordPayne82,PudritzNorman83}:
outflowing gas from accretion disks can be centrifugally accelerated along magnetic
field lines when the inclination angle of the poloidal field is above $30^\circ$ (relative
to the disk normal)\footnote{The X-wind model \citep{Shu_etal94} is also magnetocentrifugal
in nature, with the wind launched near the inner edge of the disk. We are interested in the
wind launched from radially extended region in PPDs.}.
The magnetocentrifugal wind scenario has also been extensively
explored with global simulations. Early global simulations treat the disk as a boundary
condition (i.e., razor-thin) with axisymmetry, and prescribe the rate of outflow from the
disk \citep{OuyedPudritz97a,Krasnopolsky_etal99,Krasnopolsky_etal03}. These
simulations demonstrated the robustness of the magnetocentrifugal acceleration and
collimation, and further found that the flow structure is sensitive to the prescribed rate
of mass loading from the disk, and may lead to episodic formation of jets
\citep{Ouyed_etal97,Krasnopolsky_etal99,Anderson_etal05}. More recent simulations
(most of which are two-dimensional) that do resolve the disk generally rely on artificially
prescribed and excessively large diffusion (that is unjustified) to prevent rapid magnetic
flux accumulation near the central object as mass accretes
\citep{Kato_etal02,CasseKeppens02,CasseKeppens04}, and the resulting wind
properties largely depend on the prescribed resistivity profile in the disk
\citep{Zanni_etal07}.

In reality, the wind launching process is governed by the microphysics within the disk,
with mass loading rate determined by requiring that the flow smoothly passes the slow
magnetosonic point \citep{WardleKoenigl93,Li95,OgilvieLivio01,Ogilvie12}.
It was found that launching a laminar disk wind generally requires a relatively strong
vertical background magnetic field with around equipartition strength at disk midplane
\citep{WardleKoenigl93,FerreiraPelletier95}, though weaker field is possible in the presence
of strong ambipolar diffusion \citep{Li96}. The vertical magnetic field can not be too strong,
which would make it difficult to be bent by the disk material, and would result in substantial
sub-Keplerian rotation that hinders wind-launching \citep{Shu_etal08,Ogilvie12}.
More detailed study of the wind-launching criteria and representative
solutions were presented in \citet{Konigl_etal10} and \citet{Salmeron_etal11} where
all the non-ideal MHD effects were taken into account. It {\it appears} that the MRI and
the laminar wind scenarios tend to mutually exclude each other
\citep{Salmeron_etal07}: MRI operates when the vertical magnetic field strength is
well below equipartition strength at the disk midplane, while launching an magnetocentrifugal
wind requires the vertical field strength to be around equipartition. The strong magnetic field
in the magnetocentrifugal wind scenario would drive very efficient accretion, making the disk
much more tenuous (with very small surface density) than that in the MRI scenario for a
standard accretion disk \citep{CombetFerreira08}.

To briefly summarize, MRI generally requires the background vertical field to be weak, and
has difficulty in accounting for the rapid accretion rate in PPDs, especially in the inner region
of $\sim1$AU. The alternative picture of magnetocentrifugal wind generally requires the
presence of strong background vertical field of equipartition strength that either drives accretion
that is too rapid or results in a tenuous disk, but the microphysics of the wind launching process
still requires more realistic treatment.

\subsection[]{This Work}

We conduct vertically stratified shearing-box simulations of a local patch of a PPD that include
a realistic profile of both Ohmic resistivity and AD coefficients. The diffusion coefficients are
interpolated from a pre-computed look-up table in real simulation time based on the gas
density, temperature (fixed) and ionization rate (calculated from the density profile). We have
not included the Hall effect (which is numerically more challenging and demanding), and this
is the first step towards understanding the non-ideal MHD of PPDs beyond Ohmic resistivity.
This is the first time that Ohmic resistivity and AD are simultaneously included in vertically
stratified simulations to study the gas dynamics of PPDs that incorporates the disk
microphysics in the most realistic manner. 

All our simulations except one include a weak vertical magnetic field, which is likely to be
more realistic for a local patch of a PPD. The other reason for including a vertical field is
that such a field geometry is more favorable for the MRI to operate in the presence of AD
\citep{BaiStone11}. We note that most vertically stratified shearing-box simulations to date
adopt a zero net vertical magnetic flux field geometry
(e.g., \citealp{SHGB96,Shi_etal10,Davis_etal10}) and we will show that such field geometry
would make the strength of the MRI turbulence diminishingly small due to AD.
Including net vertical magnetic flux in shearing-box simulations places strong demands on
the robustness of numerical algorithms, especially in the magnetic dominated disk corona
\citep{MillerStone00}. Recently such simulations are successfully performed by several
groups, mostly in the ideal MHD regime, and it was shown that the inclusion of net vertical
magnetic field always leads to an outflow from an MRI-turbulent disk
\citep{SuzukiInutsuka09,Fromang_etal12,BaiStone12a}. In the context of PPDs, it was
found that including a net vertical magnetic flux does not change the basic picture of
layered accretion (again, considering Ohmic resistivity only), but stronger net vertical flux
leads to stronger MRI turbulence in the active layer and reduces the vertical extent of the
dead zone, as well as stronger outflow \citep{OkuzumiHirose11}. Although these results
suggest the simultaneous existence of the MRI and magnetocentrifugal wind,
\citet{BaiStone12a} pointed that the outflow from MRI active region is unlikely to be
{\it directly} connected to a global magnetocentrifugal wind due to the MRI dynamo and
symmetry considerations. Our simulations in this paper will further address the potential
connection of the disk outflow to an magnetocentrifugal wind in the context of
PPDs, and we arrive at positive conclusions, in contrast with the ideal MHD case studied
by \citet{BaiStone12a}.


More specifically, we consider a fiducial model that corresponds to a minimum-mass solar
nebular at 1 AU, assuming solar abundance of chemical composition and $0.01\%$ (in
mass) of $0.1\mu$m sized grains for the chemistry calculation. We find that although the
initial condition is unstable to the MRI (with net vertical magnetic field much weaker than
equipartition), the disk rapidly adjusts to a new laminar configuration that is stable to the
MRI. The new laminar state is characterized by an outflow launched by the
magnetocentrifugal mechanism, and the outflow can achieve a physical wind geometry
(poloidal streamlines at the top and bottom sides of the disk bend towards the same radial
direction) by having a thin current layer where the horizontal
field flips. The magnetocentrifugal wind launched in this scenario can efficiently transport angular momentum
to account for the observed PPD accretion rate while without being too efficient to deplete
the disk (as in the conventional wind scenario). For clarity, we focus on the physical properties
of the new laminar solution and together with a parameter study all at 1 AU in this paper,
extension of the results to other disk radii will be presented in a companion paper.


This paper is organized as follows. In Section \ref{sec:method}, we describe the
numerical method and the overall setup of our simulations. Three contrasting simulations
are presented in Section \ref{sec:fiducial}, highlighting the new laminar wind solution.
In Section \ref{sec:property}, we study the physical properties of the new laminar wind
solution in detail and discuss the wind launching mechanism.
In Section \ref{sec:parameter}, we conduct a thorough parameter study to further
explore the properties of the laminar wind.
We discuss the robustness and implications of our wind solution in Section
\ref{sec:discussion}, together with the conclusion.

\section[]{Simulation Setup}\label{sec:method}

\subsection[]{Formulation}

We use the Athena MHD code \citep{Stone_etal08} to study the gas dynamics in PPDs
and perform three-dimensional numerical simulations.
We consider a local patch of a PPD and adopt the conventional shearing-box approach
\citep{GoldreichLyndenBell65}. MHD equations are written in a Cartesian coordinate
system in the corotating frame at a fiducial radius with Keplerian frequency $\Omega$:
\begin{equation}\label{eq:gascont}
\frac{\pa\rho}{\pa t}+\nabla\cdot(\rho\mb{u})=0\ ,
\end{equation}
\begin{equation}
\frac{\pa\rho\mb{u}}{\pa t}+\nabla\cdot(\rho\mb{u}^T{\mb u}
+{\sf T})=\rho\bigg[2{\mb u}\times{\mb\Omega}
+3\Omega^2x{\mb e}_x-\Omega^2z{\mb e}_z\bigg]\ ,
\label{eq:gasmotion}
\end{equation}
where ${\sf T}$ is the total stress tensor
\begin{equation}
{\sf T}=(P+B^2/2)\ {\sf I}-{\mb B}^T{\mb B}\ ,
\end{equation}
${\sf I}$ is the identity tensor, $\rho$, ${\mb u}$, $P$ are the gas density, velocity and
pressure respectively, ${\mb B}$ is the magnetic field, ${\mb e}_x,{\mb e}_y,{\mb e}_z$
are unit vectors pointing to the radial, azimuthal and vertical directions respectively, where
${\mb\Omega}$ is along the ${\mb e}_z$ direction. Note that the equations are written
in units such that magnetic permeability is 1. Vertical gravity is included to account
for density stratification. Periodic boundary conditions are used in the
azimuthal direction, while the radial boundary conditions are shearing periodic
as usual. Shearing-box source terms (Coriolis force and tidal gravities) have been readily
implemented in Athena \citep{StoneGardiner10}, which uses an orbital advection scheme
that splits the system into an advective part for the background shear flow
$-3\Omega x/2{\mb e}_y$, and a fluctuating part with velocity fluctuation ${\mb v}$:
\begin{equation}
{\mb v}\equiv{\mb u}+\frac{3}{2}\Omega x{\mb e}_y\ .
\end{equation}
The advection scheme not only accelerates the calculation by permitting larger time steps,
but also improves the accuracy by making the truncation error independent of radial location.

We use an isothermal equation of state $P=\rho c_s^2$, where $c_s$ is the isothermal
sound speed. In reality, multiple radiative processes such irradiation and scattering are
involved in PPDs, making the disk surface layer substantially hotter (e.g.,
\citealp{HiroseTurner11}). The hotter disk surface affects the properties disk wind, however,
the main problems addressed in this paper are largely magnetic: the suppression/survival
of the MRI and wind launching are largely controlled by non-ideal MHD effects, where
thermodynamics only plays a minor role. 

Being weakly ionized, PPDs are not perfectly conducting, which is reflected in the non-ideal
MHD terms in the induction equation (e.g., \citealp{Wardle07, Bai11a})
\begin{equation}
\frac{\pa{\mb B}}{\pa t}=\nabla\times({\mb u}\times{\mb B})
-\nabla\times[\eta_O{\mb J}+\eta_H({\mb J}\times{\hat{\mb B}})
+\eta_A{\mb J}_\perp]\ ,\label{eq:induction}
\end{equation}
where ${\mb J}=\nabla\times{\mb B}$ is the current density, $\hat{\mb B}$ denotes
unit vector along ${\mb B}$, subscript ``$_\perp$" denotes the vector component that is
perpendicular to ${\mb B}$, $\eta_O, \eta_H$ and $\eta_A$ are the Ohmic, Hall and the
ambipolar diffusivities. Note that ${\mb u}$ represents the velocity for the bulk of the gas
(i.e., neutrals), while tracer amount of charged species provides conductivity and gives
rise to non-ideal MHD effects.

The magnetic diffusivities depend on the number density of the charged species, and
are characterized by the dimensionless Elsasser numbers, defined as
\begin{equation}
\Lambda\equiv\frac{v_A^2}{\eta_O\Omega}\ , \qquad
Ha\equiv\frac{v_A^2}{\eta_H\Omega}\ , \qquad
Am\equiv\frac{v_A^2}{\eta_A\Omega}\ ,\label{eq:elsasser}
\end{equation}
for Ohmic, Hall and AD respectively, where
$v_A=\sqrt{B^2/\rho}$ is the Alfv\'en velocity. In the absence of grains, we have
$\Lambda\propto B^2$, $Ha\propto B$ and $Am$ being independent of $B$
(see next subsection for details). Generally speaking, self-sustained MRI turbulence
requires these Elsasser numbers to be greater than 1.

Ohmic resistivity and ambipolar diffusion (AD) have been implemented in Athena
\citep{Davis_etal10, BaiStone11}. Furthermore, we have implemented super
time-stepping that substantially accelerates the calculation, which we have described
in detail in \citet{Simon_etal13}.
Although our simulations do not include the Hall term, we do evaluate $\eta_H$ and
assess its importance in our analysis.

We use natural unit in our simulations, where $c_s=1$, $\Omega=1$. The initial density
profile is taken to be Gaussian
\begin{equation}
\rho=\rho_0\exp{(-z^2/2H^2)}\ ,
\end{equation}
with $\rho_0=1$ being the midplane gas density in code unit, and
$H\equiv c_s/\Omega=1$ is the thermal scale
height. We perform simulations with both zero and non-zero net-vertical magnetic flux.
For simulations with zero net vertical flux, the initial field is purely vertical given by
$B_z=B_1\sin{2\pi x/L_x}$, where $L_x$ is the radial size of the simulation box, and
$B_1$ parameterized by the midplane plasma $\beta_1=2\rho_0c_s^2/B_1^2$. For
simulations with net-vertical flux, we add a uniform vertical field $B_0$ (parameterized
by midplane plasma $\beta_0=2\rho_0c_s^2/B_0^2$) on top of the sinusoidally
varying component.

Physically, we consider a patch of the PPD using the minimum-mass solar nebular
(MMSN, \citealp{Weidenschilling77,Hayashi81})
disk model, with the surface density given by $\Sigma(R)=1700R_{\rm AU}^{-3/2}$g
cm$^{-2}$, and temperature given by $T=280R_{\rm AU}^{-1/2}$K, where $R_{\rm AU}$
is the radius to the central star (whose mass is fixed at $1M_{\bigodot}$) measured in
astronomical unit (AU). The mean molecular weight of the neutrals is taken to be
$\mu_n=2.34m_H$ from which the sound speed $c_s=\sqrt{kT/\mu_nm_H}=1$ km s$^{-1}$
(at 1 AU) hence other quantities can be easily evaluated. In particular, $B=1$ in our code
unit corresponds to field strength of $13.1$ Gauss. These physical scales are needed to
normalize the magnetic diffusivities (next subsection) to code unit.

We use outflow boundary condition in the vertical direction which copies the density,
velocity and magnetic fields in the boundary cells to the ghost zones, with the density
attenuated following the Gaussian profile to account for vertical gravity. In the case of
mass inflow, the vertical velocity is set to zero at the ghost zones.
A density floor of $\rho_{\rm Floor}=10^{-6}$ (in code unit) is applied to avoid numerical
difficulties at magnetic dominated (low plasma $\beta$) regions. We have checked that
horizontally averaged densities in the saturated states of all our simulations are always
well above the density floor\footnote{Except for Run OA-nb-F3 where a density floor of
$\rho_{\rm Floor}=10^{-8}$ is applied, which will be discussed in Section \ref{ssec:ADzn}.}.
Moreover, the use of outflow boundary condition no longer conserves mass in the
simulation box. We compensate for the mass loss so that a steady state can be achieved,
following the same procedure as \citet{BaiStone12a}: the density of each cell is modified
by the same proportion at the end of each time step so that the total mass in the simulation
box remains the same. For most of our runs, the mass change over the duration of
our simulations (if mass conservation were not enforced) is only a tiny fraction of the total
mass.

\subsection[]{Calculation of Magnetic Diffusivities}

The magnetic diffusivities are evaluated based on the chemistry calculation. Instead of
evolving the chemical network in real time, as done by a number of previous works
\citep{Turner_etal07,TurnerSano08,IlgnerNelson08}, we assume equilibrium chemistry
(similar to \citealp{HiroseTurner11}), because the recombination time has been
shown to be much shorter than the dynamical time scale \citep{Bai11a}. We adopt a
complex chemical reaction network \citep{IlgnerNelson06,BaiGoodman09,Bai11a}
that is based on the UMIST database \citep{Woodall_etal07}.
In our fiducial model considered in this paper, we fix the elemental composition to be
solar, with well-mixed $0.1\mu$m grains whose abundance is $\epsilon_{\rm gr}=10^{-4}$
in mass (about 0.01 solar, corresponding to substantial grain growth and settling).
We follow the same procedure and methodology described Section 3.4 in
\citet{Bai11a} to evolve the network for $10^7$ years, with further details provided in
Sections 3.2-3.5 of \citet{BaiGoodman09}. Given the chemical composition, variable
parameters of the network include gas density $\rho$, gas temperature $T$, and the
ionization rate $\xi$, which are scanned to give a complete coverage of the parameter
space relevant to PPDs. The outcome of the scan is a look-up table of magnetic
diffusivities that is read into the code so that $\eta_O$, $\eta_H$ and $\eta_A$ at each
grid cell can be evaluated by interpolation in real simulation time. Since we adopt an
isothermal equation of state, $T$ is fixed, our look-up table is essentially two-dimensional
($\rho$ and $\xi$).

The ionization rate in the disk depends on the column density to the disk surface. We
calculate the horizontally averaged vertical density profile in real simulation time, from which
a column density profile can be reconstructed, an approach similar to previous works (e.g.
\citealp{Turner_etal07}). The sources of ionization include radioactive decay, cosmic ray
and stellar X-ray ionizations, with prescriptions given in Section 3.2 of \citet{Bai11a}, with 
fixed X-ray luminosity of $10^{30}$erg s$^{-1}$ and X-ray temperature of $5$keV. In addition,
we consider the effect of far-ultraviolet (FUV) ionization. According to
\citet{PerezBeckerChiang11b}, FUV photons almost completely ionize tracer species such as
C and S and give ionization fraction of the order $f=10^{-5}-10^{-4}$ with penetration depth
$0.01-0.1$ g cm$^{-2}$ depending on the effectiveness of dust attenuation and self-shielding
\footnote{There are large uncertainties associated with the FUV ionization. For example, it
has been noted that the FUV photons may be shielded if a dusty wind is launched from the
inner disk near the dust sublimation radius \citep{BansKonigl12}. Moreover, photochemistry in
the FUV layer plays an important role in determining the molecular composition and and ion
abundances \citep{Walsh_etal12}, and particularly, the FUV penetration depth may be
overestimated in \citet{PerezBeckerChiang11b} due to the conversion of C$^+$ and S$^+$ into
molecular ions which facilitates recombination (\citealp{Adamkovics_etal11}, Al Glassgold,
private communication). Our treatment should be considered as a first
approximation that captures basic physics rather than deep into the details.}. 
For simplicity, we assume an ionization fraction of $f=2\times10^{-5}$ in the form
of carbon in the FUV layer, whose column density is chosen by default as
$\Sigma_{\rm FUV}=0.03$ g cm$^{-2}$.
Within the FUV layer, the magnetic diffusivities expressed in the form of
Elsasser numbers under the MMSN disk model are found to be
\begin{equation}
\begin{split}
Am&=\frac{\gamma\rho_i}{\Omega}\approx3.6\times10^7
\bigg(\frac{f}{10^{-5}}\bigg)\bigg(\frac{\rho}{\rho_0}\bigg)R_{\rm AU}^{-5/4}\ ,\\
Ha&=\frac{en_eB}{c\rho\Omega}\approx6.4\times10^6
\bigg(\frac{f}{10^{-5}}\bigg)\frac{1}{\sqrt{\beta_{\rm mid}}}R_{\rm AU}^{-1/8}\ ,\\
\end{split}
\end{equation}
where $\beta_{\rm mid}$ is ratio of magnetic pressure to midplane gas pressure. A smooth
transition (across about 4 grid cells) on the magnetic diffusivities from the FUV ionization
layer at $\Sigma<\Sigma_{\rm FUV}$ to the X-ray/cosmic-ray dominated ionization layer
(based on our chemistry calculations) at $\Sigma>\Sigma_{\rm FUV}$ is applied. As Ohmic
resistivity plays essentially no role in the low density region of the FUV layer, we simply
do not reset Ohmic resistivity in this layer.

We calculate the magnetic diffusivities from the number density of all charged species at
the end of the chemical evolution following Section 2 and 3.5 of \citet{Bai11a}. Note that the
Ohmic resistivity $\eta_O$ is independent of magnetic field strength $B$, while Hall and
ambipolar diffusivities do depend on $B$.
In the grain-free case, we have $\eta_H\propto B$, and $\eta_A\propto B^2$. In this case,
we can simply fit the proportional coefficients, $Q_H$ and $Q_A$ respectively, and put them
into the look-up table. In the presence of small grains, a situation studied in detail in
\citet{Bai11b}, $\eta_H$ ($\eta_A$) is proportional to $B$ ($B^2$) when $B$ is sufficiently
weak or sufficiently strong respectively, while is roughly proportional to $B^{-1}$ ($B^0$) at
some intermediate field strength. 
In this case, we include in the look-up table the two proportional coefficients $Q_{H1}, Q_{H2}$
($Q_{A1}, Q_{A2}$) at weak and strong field regimes from the fitting respectively, together with
a transition field strength $B_i$ so that 
\begin{equation}
\eta_H=\begin{cases}
Q_{H1}B\ , &
B<\sqrt{Q_{H2}/Q_{H1}}B_i\ , \\
Q_{H2}B_i^2B^{-1}\ , &
\sqrt{Q_{H2}/Q_{H1}}B_i<B<B_i\ , \\
Q_{H2}B\ , &
B>B_i\ , \\
\end{cases}\label{eq:etaA}
\end{equation}
\begin{equation}
\eta_A=\begin{cases}
Q_{A1}B^2\ , &
B<\sqrt{Q_{A2}/Q_{A1}}B_i\ , \\
Q_{A2}B_i^2\ , &
\sqrt{Q_{A2}/Q_{A1}}B_i<B<B_i\ , \\
Q_{A2}B^2\ , &
B>B_i\ , \\
\end{cases}\label{eq:etaA}
\end{equation}
By comparing with Figure 1 of \citet{Bai11b}, we see that the transition field strength
$B_i$ corresponds to the situation that the ion gyro-frequency equals its collision
frequency with the neutrals (or the ion Hall parameter equals one), which can be
directly calculated given the gas density.

Being diffusive processes, large Ohmic resistivity near the disk midplane and strong
AD in the tenuous disk corona would significantly limit the code efficiency even super
time-stepping is used to accelerate the calculation. In the Ohmic regime, the diffusive
time step scales as $\Delta^2/\eta_O$ where $\Delta$ is the minimum grid spacing.
In practice, we add a cap to the Ohmic resistivity so that in code unit $\eta_O\leq10$.
Since the magnetic field strength near the disk midplane never reaches equipartition
in all our simulations, the Elsasser number at the disk midplane is always smaller than
$0.1$, well below the threshold value $1$. Also, this cap value of $\eta_O$ makes the
diffusion time scale much smaller than the dynamical time scale, hence captures the
basic effect of strong diffusion at disk midplane even if resistivity is much higher in
reality. Note that our resistivity cap is much larger than most previous works (thanks to
the use of super time-stepping), where the cap was of the order $0.01$ in natural unit
(e.g., \citealp{FlemingStone03,OkuzumiHirose11}).
In the AD regime, the time step scales as $\Delta^2\cdot Am\cdot\beta$ ( $\beta$ is the
local ratio of gas to magnetic pressure). In the same spirit, we apply a floor to
$Am$ so that $Am\cdot\beta\geq0.1$ in every grid cell.
This floor value of $Am$ is again sufficiently small so that it does not make the otherwise
stable field configuration unstable to the MRI (see Figure 16 of \citealp{BaiStone11}),
and it retains the effect of strong diffusion.

\section[]{Fiducial Simulations and Results}\label{sec:fiducial}

In this section, we present three benchmark simulations with very similar
initial setup but evolve into dramatically different states. All three simulations
are three-dimensional (3D) shearing-box with vertical stratifications, located
at 1 AU in a MMSN disk, adopting a chemistry model with well-mixed
$0.1\mu$m grains with abundance of $10^{-4}$ ($1\%$ solar), and a FUV
column density $\Sigma_{\rm FUV}=0.03$ g cm$^{-2}$. All simulations are
run for about 150 orbits ($900\Omega^{-1}$).

In the first simulation (Run O-b5), only Ohmic resistivity is included, which
aims at modeling the conventional picture of layered accretion. We have also
included a net vertical magnetic flux of $\beta_0=10^5$. A sinusoidally varying
vertical field of $\beta_1=\beta_0/4$ is also added as initial condition to
avoid the strong channel flows (but has no effect on the saturated state of the
system). The simulation box size is $4H\times8H\times16H$ in the radial ($x$),
azimuthal ($y$) and vertical ($z$) dimensions respectively, with a computational
grid of $96\times96\times384$ cells. Our simulations have relatively high
resolution in $x$ and $z$ (24 cells per $H$, or 34 cells if one defines the scale
height to be $\sqrt{2}H$, as in a number of works) to properly resolve the MRI
turbulence \citep{Davis_etal10,Sorathia_etal12,BaiStone11}, and have relatively
large simulation box to capture the mesoscale structures of the MRI turbulence
\citep{Simon_etal12a}. 

In the second simulation (Run OA-nobz), both Ohmic resistivity and AD are
included, and we adopt a zero net vertical magnetic flux configuration with
$\beta_1=1600$. Again, the choice of $\beta_1$ has no effect on the
saturated state of the system. After experimenting with several initial setups,
we find that due to the extremely weak level of turbulence, the gas near the
disk surface has very limited magnetic support hence drops very rapidly, and
we must reduce the box height to $12H$ (and $288$ cells) so that the gas
density at vertical boundaries is not too small which would severely reduce
the numerical time step. Furthermore, we started the simulation with density
floor $\rho_{\rm floor}=10^{-6}$, but then found that the density at vertical
boundaries is always at the level of the density floor. We thereby keep reducing
the density floor until $\rho_{\rm floor}=10^{-8}$ when we no longer find
artificial features near the vertical boundaries.

Finally, we conduct a contrasting simulation from the above two cases
(Run OA-b5). The initial setup of the simulation is exactly the same as in the
first simulation (Run O-b5) except that AD is included. 

\begin{figure}
    \centering
    \includegraphics[width=90mm]{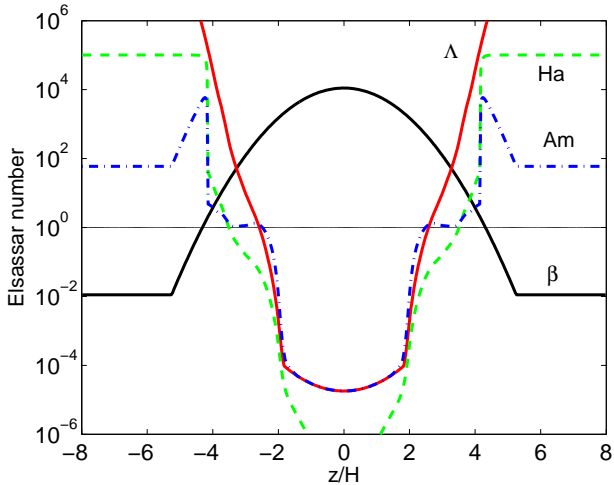}
  \caption{Initial Elsasser number profile for Ohmic resistivity (red solid), Hall diffusivity
  (green dashed) and ambipolar diffusivity (blue dash-dotted) for our fiducial runs
  O-b5 and OA-b5 . Note the cap on $Am$ and $\Lambda$ around the disk midplane
  and on $Am$ beyond about $\pm5H$. Also shown is the initial profile for plasma
  $\beta$ (black solid). Note that we show the Hall Elsasser number ($Ha$) while it
  is not included in the simulations.}\label{fig:ini_els}
\end{figure}

Figure \ref{fig:ini_els} illustrates the {\it initial} profile of the Ohmic, Hall and
ambipolar Elsasser numbers for our runs O-b5 and OA-b5 (similar but not
exactly the same for run OA-nobz due to the different magnetic configuration).
Also shown is the initial profile of plasma $\beta$. From the disk midplane to
surface, the dominant non-ideal MHD effects are Ohmic resistivity, Hall effect
and AD respectively for the initial field configuration, and MRI unstable region
is located at around $z=\pm3H$ where all Elsasser numbers are greater than 1
and plasma $\beta$ is well above 1. The initial FUV ionization front is located at
about $z=\pm4H$. The density floor of $\rho_{\rm floor}=10^{-6}$ is applied to
regions beyond $z=\pm5H$, hence the $Am$ and $\beta$ curves flattens out
(this artifact will disappear as the system evolves).

The most common accretion diagnostics is the $R\phi$ component of the
Reynolds and Maxwell stresses, which measures the local rate of radial
transport of angular momentum
\begin{equation}
T_{R\phi}=T_{R\phi}^{\rm Rey}+T_{R\phi}^{\rm Max}
=\overline{\rho v_xv_y}-\overline{B_xB_y}\ ,
\end{equation}
where the over bar indicates horizontal averaging. The total rate of radial
angular momentum transport is characterized by the $\alpha$ parameter
\citep{ShakuraSunyaev73}
\begin{equation}
\alpha\equiv\frac{\int T_{R\phi}dz}{c_s^2\int \rho dz}\ .\label{eq:alpha_def}
\end{equation}

\begin{figure*}
    \centering
    \includegraphics[width=180mm]{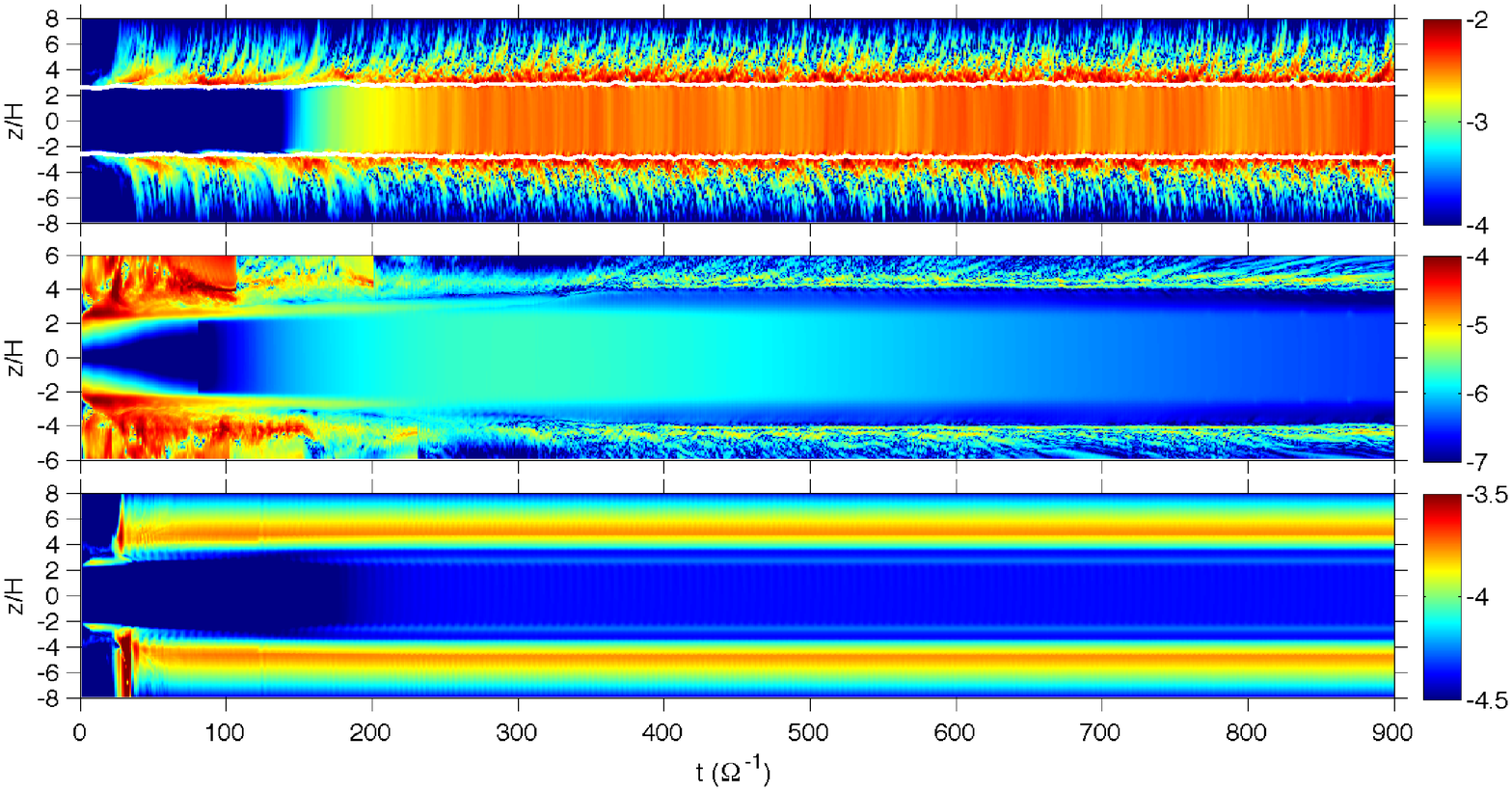}
     \caption{The space-time plot for the (horizontally averaged) vertical profiles of the
     Maxwell stress ${-\overline{B_xB_y}}$ in the three fiducial runs O-b5 (top), OA-nobz
     (middle) and OA-b5 (bottom). The colors are in logarithmic scales (in code unit).
     White contours in the upper panel correspond
     to vertical Elsasser number $\Lambda_z\equiv v_{Az}^2/\eta_O\Omega=1$. The
     discontinuous transitions in the middle panel correspond to the sudden reduction
     of the density floor which attains the final value of $\rho_{\rm floor}=10^{-8}$ at
     $t=230\Omega^{-1}$.}\label{fig:history}
\end{figure*}

In Figure \ref{fig:history} we show the time evolution of the horizontally
averaged Maxwell stress $-\overline{B_xB_y}$ for the three fiducial
simulations. Albeit for similar initial setup, the three runs show very distinctive
characteristics as the systems evolve into saturated/steady states. The
results from the three simulations will be analyzed in detail in the following
three subsections.

For clarity, we further provide a list of all our fiducial simulation runs and their
parameters in Table \ref{tab:fidruns}. In particular, we introduce the letter ``S"
for runs with very small horizontal domain, where the simulations are essential
one-dimensional, and letter ``E" for runs with enforced even-$z$ symmetry
(see Section \ref{sec:property}). We use the label ``bn" to denote plasma
$\beta_0=10^n$ for the vertical background field.
All other simulations are run for about 200 orbits ($1200\Omega^{-1}$).

\begin{table}
\caption{Summary of All Fiducial Simulations.}\label{tab:fidruns}
\begin{center}
\begin{tabular}{ccccc}\hline\hline
 Run & Diffusion & $\beta_0$ 
 & Box Size & Section\\\hline
O-b5 & Ohm & $10^5$ & $4H\times8H\times16H$ & 3.1\\
OA-nobz & Ohm, AD & $\infty$ & $4H\times8H\times12H$ & 3.2 \\
OA-b5 & Ohm,AD & $10^5$ & $4H\times8H\times16H$ & 3.3, 4\\
S-OA-b5 & Ohm,AD & $10^5$ & $H/8\times H/4\times16H$ & 4.4\\
E-OA-b5 & Ohm,AD & $10^5$ & $4H\times8H\times8H$ & 4.4.1\\
S-OA-b5-12H & Ohm,AD & $10^5$ & $H/8\times H/4\times12H$ & 4.5\\
S-OA-b5-20H & Ohm,AD & $10^5$ & $H/8\times H/4\times20H$ & 4.5\\
S-OA-b5-24H & Ohm,AD & $10^5$ & $H/8\times H/4\times24H$ & 4.5\\
\hline\hline
\end{tabular}
\end{center}
MMSN disk model at 1AU, X-ray luminosity of $L_X=10^{30}$ ergs s$^{-1}$
and temperature $T_X=5$keV, well mixed $0.1\mu$m grains with abundance
of $\epsilon_{\rm gr}=10^{-4}$, penetration depth of $0.03$g cm$^{-2}$ for
the FUV ionization are assumed for all runs.
\end{table}


\subsection[]{The Ohmic-Resistivity-Only Run}\label{ssec:Ohm}

\begin{figure}[t]
    \centering
    \includegraphics[width=85mm]{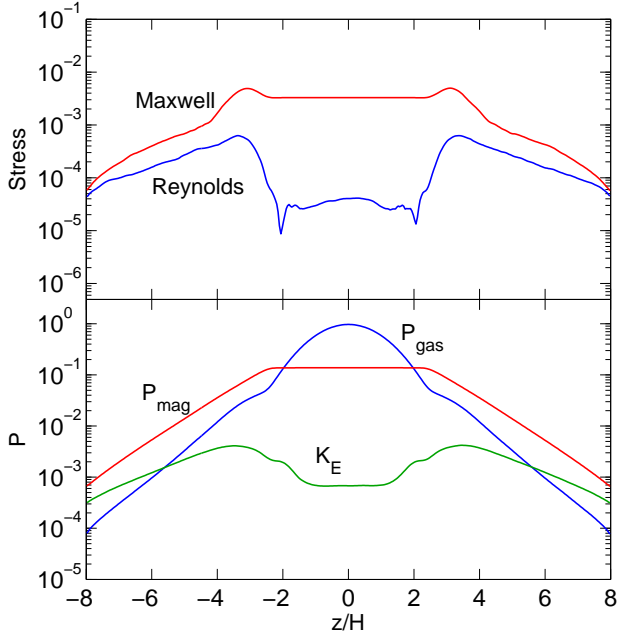}
  \caption{Vertical profiles of the Maxwell stress and Reynolds stress (upper panel), as well
  as gas/magnetic pressure and kinetic energy (lower panel) in the fiducial run with only Ohmic
  resistivity (O-b5).}\label{fig:FidO-prof}
\end{figure}

Run O-b5 quickly develops into turbulence. From the upper panel of Figure \ref{fig:history},
the separation between the highly turbulent active layer and the more or less quiescent
midplane region is clearly seen. We further show the time and horizontally averaged
vertical profiles of density, magnetic pressure, Maxwell and Reynolds stresses in
Figure \ref{fig:FidO-prof}. The time averages are performed from $\Omega t=450$ onward.

The MRI turbulence generates strong magnetic field that buoyantly rises, and the disk
surface quickly forms a strongly magnetically dominated corona beyond $\pm2H$
\citep{MillerStone00}, as shown in the bottom panel of Figure \ref{fig:FidO-prof}. The gas
density in the corona follows an exponential profile due to strong magnetic
support, instead decreasing with height as an Gaussian in the hydrostatic case. The
velocity in the coronal regions is highly supersonic, with turbulent kinetic energy exceeding
the gas pressure beyond $\pm5H$. A strong outflow is launched from the active layer of
the disk as has been studied by \citet{Suzuki_etal10}, who also found that the mass
outflow rate scales roughly linearly with the vertical net magnetic flux. In our simulation,
the time and horizontally averaged mass outflow rate $\bar{\rho v_z}$ is found to be about
$3.7\times10^{-4}\rho_0c_s$ from each side of the box, comparable with the
measurements in \citet{Suzuki_etal10} with the same $\beta_0$. We note that the mass
outflow rate is not a well-characterized quantity in shearing-box \citep{BaiStone12a}, these
measurements should only be taken as a reference.

The disk midplane is too resistive to become MRI active, with the boundary of the active layer
well described by the Elsasser number criterion: $\Lambda_z\equiv v_{Az}^2/\eta_O\Omega=1$,
where $v_{Az}$ is the vertical component of the Alfv\'en velocity. The Maxwell stress remains
very small at the midplane for the first 20 orbits. However, the midplane magnetic field is then
gradually amplified, while the flow in the midplane remains more or less laminar. We note that
the Maxwell stress in the midplane is even larger than most regions in the active layer, which
contrasts with some previous simulations with a similar setup, where the Maxwell stress
becomes very small near the midplane \citep{Suzuki_etal10,HiroseTurner11,OkuzumiHirose11}.
The main reason for the difference, while somewhat counterintuitive, lies in the usage of a much
larger resistivity cap in our simulations\footnote{See last paragraph of Section 2. In fact, we start
the simulation with a smaller resistivity cap of $\eta_O=1$ in code unit. At $t=120\Omega^{-1}$,
the cap is raised to its final value $\eta_O=10$, and the Maxwell stress at disk midplane rises
shortly afterwards.}. The large resistivity at the disk midplane in our simulations strongly
suppresses electric current, leaving the horizontal magnetic field to be almost constant across
the midplane (see the bottom panel of Figure \ref{fig:FidO-prof}). Therefore, the midplane field
strength is largely set by the field strength in the active layer. Note that although this is
phenomenologically similar to the ``undead" zone proposed by \citet{TurnerSano08}, it is
conceptually very different. We note that it is essential for the resistivity cap to be large enough
so that the gas and magnetic field become decoupled at the cap value, and we have further
tested that the simulation results are independent of the resistivity cap once its value is greater
than $3$ or so.

By contrast, the Reynolds stress $\overline{\rho v_xv_y}$ as well as kinetic energy clearly
have a large dip in the undead region between $z=\pm2H$. There is still random motion
near the midplane due to the sound waves launched from the base of the active layer
\citep{FlemingStone03,OishiMacLow09}, though the velocity amplitude is at least one order
of magnitude smaller than that in the active layer.

Integrating the profiles of the Maxwell and Reynolds stresses using (\ref{eq:alpha_def}),
we obtain the Shakura-Sunyaev parameter $\alpha_{\rm Max}\approx1.3\times10^{-2}$,
and $\alpha_{\rm Rey}\approx1.2\times10^{-3}$. Assuming steady state accretion, one can
express the accretion rate for a MMSN disk model as
\begin{equation}
\dot{M}=\frac{2\pi\alpha\Sigma c_s^2}{\Omega}\approx
8.2\times10^{-6}\alpha R_{\rm AU}^{-1/2}M_{\bigodot}\ {\rm yr}^{-1}\ ,\label{eq:mdot1}
\end{equation}
Therefore, in the absence of AD, and at the fiducial location $R=1$ AU, we obtain an
accretion rate of $1.2\times10^{-7}M_{\bigodot}$ yr$^{-1}$, large enough to
account for the observed accretion rates in most T-Tauri stars.

Given the usage of more realistic resistivity profiles and chemistry models, as well as the
much larger resistivity cap enabled by the super time-stepping technique, our run with
pure Ohmic-resistivity deserves more discussion on its own right. Nevertheless, we only
consider our run O-b5 as a test and reference case since our main point of interest is the
effect of AD on the gas dynamics of PPDs, which makes a dramatic difference from the
conventional picture of PPD accretion.

\subsection[]{Zero Net Vertical Flux Run with both Ohmic Resistivity and AD}\label{ssec:ADzn}

\begin{figure}
    \centering
    \includegraphics[width=85mm]{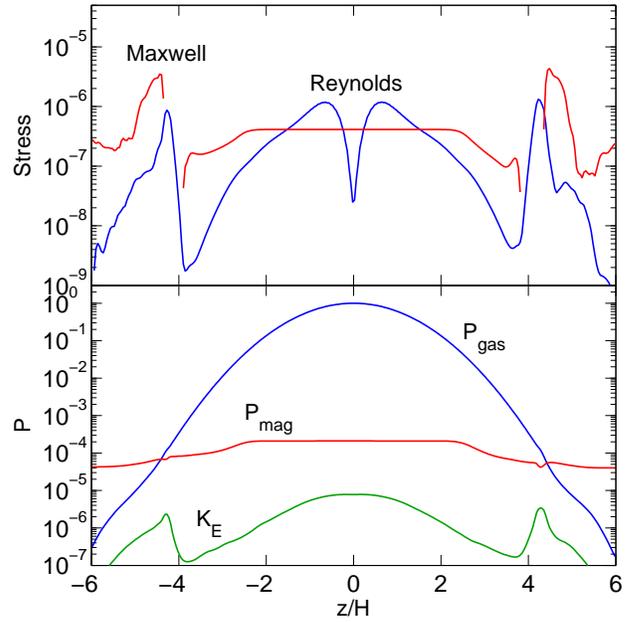}
  \caption{Same as Figure \ref{fig:FidO-prof}, but for zero net vertical flux run with both
  Ohmic resistivity and AD (OA-nobz).Note the different scales.}\label{fig:FidOAnobz-prof}
\end{figure}

Setting the initial density floor of $\rho_{\rm floor}=10^{-6}$, we find that the MRI turbulence
sets in at the beginning, then the simulation gets stuck with relatively strong
magnetic field (dominantly toroidal) accumulating near the vertical boundaries with
relatively little activities. The flux does not escape and provides very little magnetic
support due to the flat profile, and the density near the vertical boundaries stays at the
value of $\rho_{\rm floor}$. This situation implies that the gas near vertical boundaries
tends to fall back towards disk midplane, but this is prevented by the outflow boundary
condition. Therefore, we gradually lower the density floor over the course of the
simulation and find that the phenomena of artificial magnetic flux accumulation and
density cutoff at $\rho_{\rm floor}$ disappears when $\rho_{\rm floor}$ is
brought down to $10^{-8}$, which is achieved at $t=230\Omega^{-1}$. The whole process
is reflected in Figure \ref{fig:history}. From this time onward, the system evolves smoothly
and no artifacts are seen from the vertical boundaries. Also note that setting such small
density floor of $10^{-8}$ at the beginning would lead to dramatically small simulation
timestep which is not quite realistic, hence gradual reduction of $\rho_{\rm floor}$ is
necessary, although one has to be patient to get rid of the artifacts.

We see from Figure \ref{fig:history} that in the saturated state, the disk only has
extremely weak level of MRI turbulence in a very thin layer between about $z\sim4-5$H.
Note that the FUV ionization front is located at $z=\pm4H$, beyond which we find
$Am\gtrsim50$ and the gas behaves almost as in the ideal MHD regime (since
density profile is still largely Gaussian, one can still use Figure \ref{fig:ini_els} as a
reference for the profile of $Am$), hence the MRI operates in this region. Below
$z=4H$, the value of $Am$ is of order unity or smaller, and there is no evidence
of turbulent activity. This is consistent with unstratified simulation results of
\citet{BaiStone11}, where it was found that the MRI is suppressed for $Am\lesssim1$
in the absence of net vertical magnetic flux.

There is some residual Maxwell stress around the disk midplane as a result of
initial conditions that slowly diminishes over the course of the simulation. We
extract the profiles of various physical quantities and show them in Figure
\ref{fig:FidOAnobz-prof}, where the time average is performed between
$\Omega t=750$ and $900$. The MRI-active region exhibits as bumps in the stress
plot, as well as the bumps in the kinetic energy plot. The magnetic field strength
roughly stays constant through the entire disk, which is likely set by the turbulent
activities in the active zone: the value of plasma $\beta$ (ratio of gas pressure to
magnetic pressure) reaches order unity within the turbulent layer. Moreover, we
find that the magnetic field is predominantly toroidal in the entire disk.

From the simulation, we find the Shakura-Sunyaev parameter using (\ref{eq:alpha_def})
to be $\alpha_{\rm Max}\approx1.7\times10^{-6}$, and
$\alpha_{\rm Rey}\approx1.3\times10^{-6}$. For such small level of stress, we find that
the resulting accretion rate is only about $2.5\times10^{-11}M_{\bigodot}$ yr$^{-1}$,
which is three orders of magnitude too small compared with observations.

We can compare this result with optimistic predictions of the MRI-driven accretion rate
using the semi-analytical framework of \citet{Bai11a}. Using this framework, we first
extract the density profiles and the profiles of $Am$ and $\eta_O$. Assuming constant
magnetic field strength across the MRI-active layer, the MRI-driven accretion rate can
be expressed as (Equation (28) of \citealp{Bai11a})
\begin{equation}
\dot{M}\approx\frac{c_s}{8\Omega^2}\int_{\rm active}B^2\frac{dz}{H}\ ,
\end{equation}
where the integral is performed in regions where the MRI is permitted based on
the criteria (20) of \citet{Bai11a}. We scan the field strength to maximize $\dot{M}$,
which gives $6.6\times10^{-10}M_{\bigodot}$ yr$^{-1}$ as an upper limit of the
MRI-driven accretion rate. We see that this value is a factor of more than $20$ larger
than our simulation result. The main reason that our simulation yields an accretion
rate that is significantly smaller than theoretical expectations is that the field
geometry is not optimal: in both the ideal MHD (e.g., \citealp{HGB95,BaiStone12a})
and non-ideal MHD (e.g., \citealp{Fleming_etal00,BaiStone11}), one finds stronger
turbulence in the presence of a net vertical magnetic flux, and the optimistic
estimate can possibly be achieved only when a optimal field geometry is realized.

This simulation has also demonstrated the physical significance of AD which
drastically reduces the accretion rate compared with the Ohmic-only case. Although
such a comparison may be unfair since our Ohmic-only simulation contains net
vertical magnetic flux, Ohmic-only simulations with similar setup and zero net vertical
flux generally yield total stress level that is only slightly smaller than ours
(e.g., \citealp{Turner_etal07,IlgnerNelson08}). Clearly, with AD taken into account,
zero net vertical magnetic field geometry is far from
being capable of driving rapid accretion in the inner region of PPDs.

\subsection[]{Net Vertical Flux Run with Both Ohmic Resistivity and AD}

\begin{figure*}
    \centering
    \includegraphics[width=160mm]{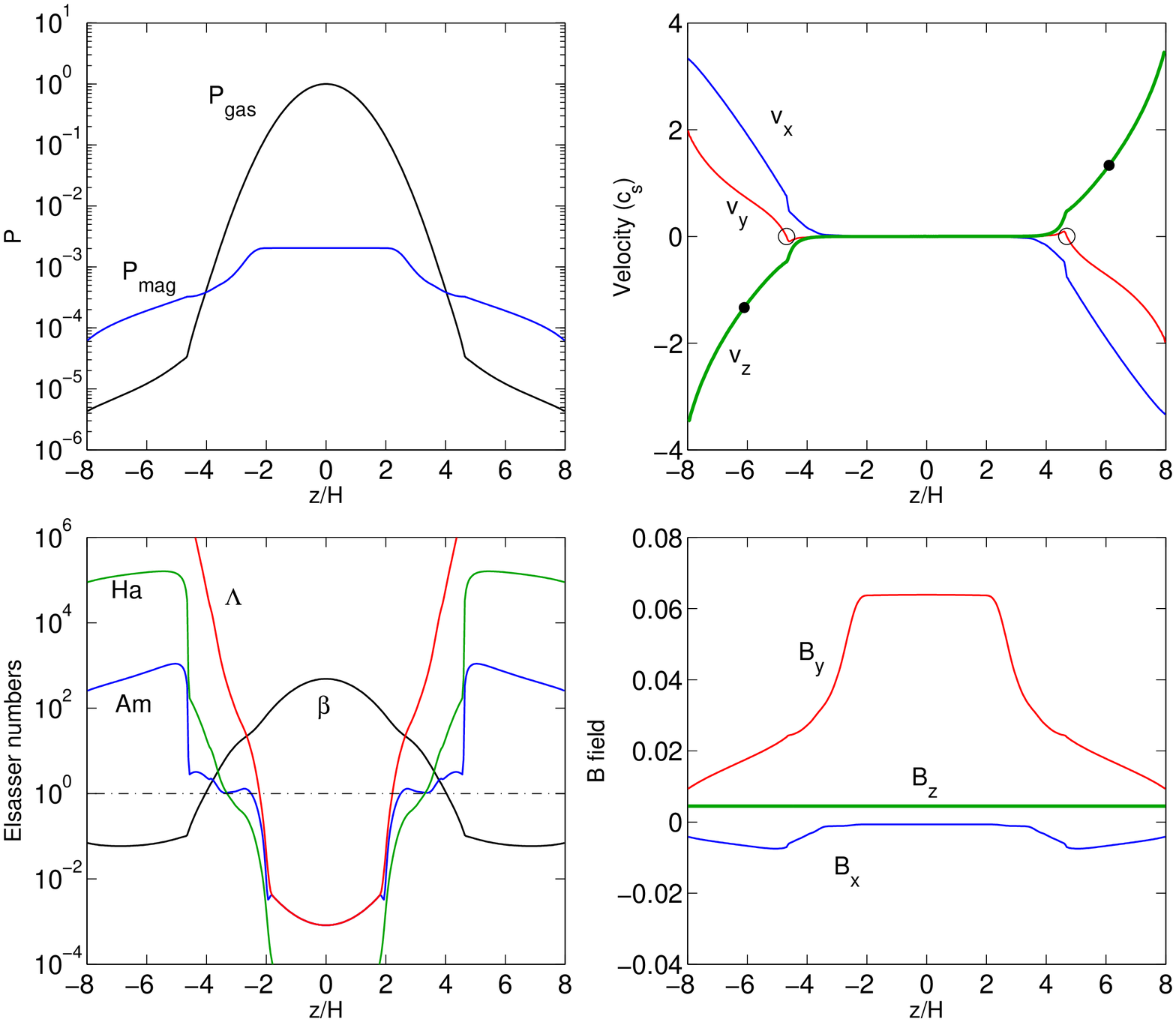}
  \caption{Vertical profiles of various quantities in the fiducial run with both Ohmic resistivity
  and AD and net vertical magnetic flux of $\beta_0=10^5$ (OA-b5). Upper left: gas pressure
  and magnetic pressure. Lower left: Elsasser numbers for Ohmic resistivity ($\Lambda$),
  Hall term ($Ha$) and AD ($Am$), together with the plasma $\beta=P_{\rm gas}/P_{\rm mag}$.
  Note that we evaluate $Ha$ while the Hall effect is not included in the simulations.
  Upper right: three components of gas velocity, where the bold green curve is for the vertical
  velocity. The Alfv\'en points are indicated as black dots, while open circles mark the base of
  the outflow. Lower right: three components of the magnetic field, where the bold green curve
  is for vertical field.}\label{fig:FidOA-prof}
\end{figure*}

The inclusion of net vertical magnetic flux in our fiducial run with both Ohmic resistivity
and AD (OA-b5) makes the field geometry more favorable for the MRI as we originally
expect. The bottom panel of Figure \ref{fig:history} illustrates the time evolution of
Maxwell stress. The initial field configuration is unstable to the MRI, which gives rise to
channel-flow like behaviors and initiates some turbulent activities in the surface layer.
However, we find that quite surprisingly, the system then quickly relaxes into a
non-turbulent state in about 10 orbits with the MRI suppressed completely. The
laminar configuration is then maintained for the remaining of the simulation
time\footnote{To justify the validity of this result, particularly that it is not due to an
unrealistic initial condition, we restart from the end of run O-b5 with AD turned on and
find that also in about 10 orbits of time, the system settles to the same laminar state.}.

As the system settles to a completely laminar state, we extract the exact
vertical profiles of various physical quantities, as shown in Figure \ref{fig:FidOA-prof}.
The magnetic field is strongest within about $\pm2.5H$ of the disk midplane, where the field
is essentially constant due to the large resistivity and AD. There is no distinction between
active layer and dead zone as the entire disk is laminar. The disk becomes magnetically
dominated beyond $z=\pm4H$. The FUV ionization front is located at about $z=\pm4.5H$
as seen in the sharp increase of $Am$ and $Ha$ profiles, which also corresponds to the
point where the gas density starts to deviate from Gaussian and follow an exponential
profile.

Being a 3D time-dependent simulation, the fact that the system reaches a steady laminar
state suggests the stability (particularly, against the MRI) of the configuration.
This stability can be qualitatively understood using the criterion based on MRI simulations
that include individual non-ideal MHD effects. In regions near the disk midplane where
Ohmic resistivity dominates, the Ohmic Elsasser number is below one within about
$z=\pm2H$, too small for the MRI to operate \citep{Turner_etal07}. Beyond this region
where AD is the dominant non-ideal MHD effect, which requires weak magnetic field for
the MRI to operate, the magnetic field is too strong, as judged from Figure 16 of
\citet{BaiStone11}. Even the FUV ionization increases $Am$ substantially beyond
$z=\pm4.5H$, the disk has already become magnetically dominated ($\beta<1$) in these
regions. The suppression of the MRI can be understood as a result of magnetic field
amplification in the surface layer during the initial growth from the MRI-unstable
configuration. The MRI is quenched once the field becomes too strong for it to operate
because of AD.

The fact that MRI is suppressed in the disk surface layer implies that the hypothesis of
\citet{Bai11a} does not always hold, where it was assumed that the magnetic field can
be amplified by the MRI to maximize the efficiency of the MRI. In other words, we see
that the magnetic field is amplified to much greater extent that the MRI is suppressed.
Nevertheless, this result is not inconsistent with the
framework of \citet{Bai11a} since it only estimates the upper limit of MRI-driven accretion
rate, and complete suppression of the MRI simply means the rate is zero.

The most prominent feature of the laminar state is a strong outflow that leaves the vertical
boundaries. All three components of the velocities exceed the sound speed relative to the
background Keplerian flow, with $v_z$ reaches about $3c_s$ at the
vertical boundaries, and an outflow mass loss rate of $\rho v_z=1.5\times10^{-5}\rho_0c_s$
from each side of the simulation box. The Alfv\'en critical point, given by
$v_z=v_{Az}=B_z/\sqrt{\rho}$, is contained within our simulation box and is indicated in
the upper right panel of Figure \ref{fig:FidOA-prof}. Moreover, according to the conventional
definition \citep{WardleKoenigl93}, we define the base of the outflow at the location where
the azimuthal velocity starts to become super-Keplerian, and the it will become more
obvious later about this definition (see Figure \ref{fig:radstress}) .

Our fiducial run OA-b5 suggests that the structure of the disk has a pure
one-dimensional (1D) profile. To verify this result, we perform a new simulation
(run S-OA-b5) with everything kept the same except that the horizontal domain is
reduced to $0.125H\times0.25H$ resolved by $4\times4$ cells. We will refer to this
type of runs as ``quasi-1D" simulations. Such small box is obviously too small to
study the MRI, but we
find that except for different initial evolution of the original MRI-unstable configuration,
the system relaxes to exactly the same laminar state with a strong outflow as in the
fiducial run OA-b5.  Moreover, we have further checked that although the initial evolution
involves horizontal variations,  such variations vanish
as the system relaxes to the final steady state\footnote{However, if we start from a pure
one-dimensional profile, the system will remain one-dimensional but does not relax to a
steady state.}. Therefore, we conclude that the inclusion of AD makes the disk structure
purely 1D.

In sum, we have seen that the three fiducial simulations differ with each other in only
one piece of physics, while show dramatically different characteristics. These
results best demonstrate the importance of including AD in the study of gas dynamics
in PPDs, as well as the significant role played by the magnetic field geometry. They
strongly suggest that the MRI operates extremely ineffectively in the inner region of
PPDs. In the mean time, the new laminar solution from our run OA-b5 strongly points to
the magnetocentrifugal wind as the most promising driving mechanism for PPD
accretion. The richness of the new findings deserves detailed study, and we devote the
next section to address the nature of the new laminar wind solution. Moreover, the 1D
nature of the new solution allows us to perform the simulations using very small
horizontal domains which tremendously reduces the computational cost.

\section[]{Nature of the Wind Solution}\label{sec:property}

In this section, we do detailed analysis of our new laminar wind solution from run
OA-b5. Table \ref{tab:windresults} summarizes the main physical properties of our
fiducial laminar wind run together with a number of companion runs as they will be
elaborated in the text.

\begin{table*}
\caption{Wind Properties from All Fiducial Simulations.}\label{tab:windresults}
\begin{center}
\begin{tabular}{ccccccccc}\hline\hline
 Run & $\alpha_{\rm Max}$ & $T_{z\phi}^{\rm Max}$ & $\dot{M}_w$
 & $z_b$ & $z_{\rm A}$ & $P_{\rm sh}$ & $\dot{E}_P$ & $\dot{E}_K$\\\hline

OA-b5 & $2.27\times10^{-4}$ & $1.07\times10^{-4}$ & $3.06\times10^{-5}$
& 4.69 & 6.10 & $1.90\times10^{-3}$ & $7.66\times10^{-4}$ & $4.38\times10^{-4}$ \\

S-OA-b5 & $2.42\times10^{-4}$ & $1.06\times10^{-4}$ & $2.86\times10^{-5}$
& 4.63 & 6.22 & $2.02\times10^{-3}$ & $8.05\times10^{-4}$ & $4.03\times10^{-4}$ \\

E-OA-b5 & $1.38\times10^{-4}$ & $1.05\times10^{-4}$ & $2.72\times10^{-5}$
& 4.56 & 6.23 & $1.61\times10^{-3}$ & $7.30\times10^{-4}$ & $4.21\times10^{-4}$ \\

S-OA-b5-12H & $2.09\times10^{-4}$ & $9.67\times10^{-5}$ & $4.13\times10^{-5}$
& 4.45 & 5.33 & $1.28\times10^{-3}$ & $4.63\times10^{-4}$ & $2.28\times10^{-4}$\\

S-OA-b5-20H & $2.60\times10^{-4}$ & $1.13\times10^{-4}$ & $2.23\times10^{-5}$
& 4.70 & 7.12 & $2.77\times10^{-3}$ & $1.18\times10^{-3}$ & $5.32\times10^{-4}$ \\

S-OA-b5-24H & $2.58\times10^{-4}$ & $1.18\times10^{-4}$ & $2.02\times10^{-5}$
& 4.83 & 7.67 & $3.25\times10^{-3}$ & $1.14\times10^{-3}$ & $8.55\times10^{-4}$ \\
\hline\hline
\end{tabular}
\end{center}
$\alpha_{\rm Max}$: normalized Maxwell stress in the disk zone;
$T_{z\phi}^{\rm Max}$: wind stress at wind base; $\dot{M}_w$: total mass loss rate
(from both sides of the disk); $z_b$: location of the wind base; $z_A$: location of
the Alfv\'en point; $P_{\rm sh}$: rate of work done at the shearing-box boundaries;
$E_P$: energy loss rate due to Poynting flux; $E_K$: kinetic part of the energy loss
rate. All quantities are in natural unit ($\rho_0=c_s=\Omega=1$).
\end{table*}

\subsection[]{Field Line Geometry and Wind Launching}\label{ssec:launching}

\begin{figure}
    \centering
    \includegraphics[width=90mm]{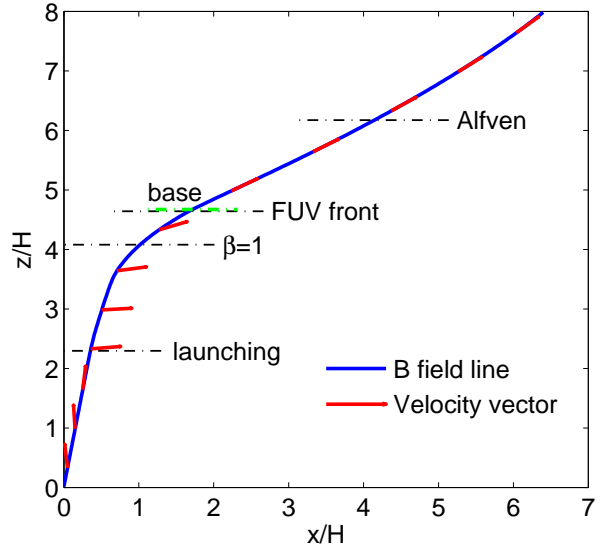}
  \caption{The poloidal field line geometry in our fiducial run OA-b5 (blue solid line).
  Overplotted are the unit vectors of the poloidal gas velocity (red arrows). The location
  of the wind launching point, the plasma $\beta=1$ point, the FUV ionization front and
  the Alfv\'en point are indicated (black dash-dotted).
  Also marked is the location at the base of the wind (green dashed).}\label{fig:fieldline}
\end{figure}

We first consider the geometry of poloidal magnetic field lines, from which much
insight can be gained on the wind launching mechanism. Using the magnetic field
vectors from our run OA-b5, we integrate a poloidal field line from the disk
midplane all the way to the vertical boundary of our simulation box, and show the
results in Figure \ref{fig:fieldline}. In the mean time, we overplot the direction
of velocity vectors as red arrows.

The field line is straight within about $\pm2.5H$ from the midplane due to the extremely
large resistivity, where the gas and magnetic field are essentially
decoupled as we discussed in the previous section. The field lines start to bend once the
gas become partially coupled to the magnetic field, characterized by Ohmic as well as AD
Elsasser numbers $\Lambda$ and $Am$ exceeding unity, which occurs at $z\simeq\pm2.3H$,
as can be seen from the bottom left panel of Figure \ref{fig:FidOA-prof}. We label this
point as the launching point in Figure \ref{fig:fieldline}. Beyond this point, the azimuthal
magnetic field decreases rapidly with height, creating large current density in the radial
direction. Together with the vertical field, we obtain the following force balance equation in
the azimuthal direction
\begin{equation}
B_z\frac{dB_y}{dz}-\frac{1}{2}\rho\Omega v_x=0\ ,\label{eq:balance}
\end{equation}
which states that the Lorentz force is balanced by the Coriolis force. This explains the
increase of radial velocity in the upper right panel of Figure \ref{fig:FidOA-prof} beyond
about $\pm2.5H$, as well as the direction of the arrows in Figure \ref{fig:fieldline} at
the same locations. In this region, AD is the dominant non-ideal MHD effect. The radial
motion of the gas makes the velocity vector deviate from the direction of the magnetic
field, which drags and bends the magnetic field lines towards the same direction via the
ion-neutral drag. Correspondingly, $|B_x|$ increases in strength.

The field lines tend to become straight again as gas density decreases and the flow
becomes magnetically dominated with total plasma $\beta<1$. Beyond this point, the
gas only has limited effect on the field line geometry. Eventually, at the base of the
wind (located at $z_b=\pm4.7H$), defined as the location where the azimuthal velocity
exceeds the Keplerian velocity, the magnetocentrifugal acceleration starts to operate.
Throughout this paper, we define the region between $z=\pm z_b$ as the disk zone,
while regions beyond as the wind zone.

We have also indicated the location of the FUV ionization front, beyond which the gas
behaves more or less as ideal MHD due to the large ionization fraction. We see that the
poloidal velocity of the gas is aligned with the poloidal magnetic field beyond the FUV
front, as expected for an ideal MHD wind. Below the FUV front, gas velocity deviates
from the direction of the magnetic fields as a result of AD and Ohmic resistivity. In this
fiducial run with $\beta_0=10^5$, the location of the FUV front overlaps with the base
of the wind, which we note is just a coincidence. The importance of the FUV ionization
will further be discussed in Section \ref{ssec:fuv}.

\subsection[]{Conservation Laws}

A laminar magnetized disk wind is characterized by the conservation of mass,
angular momentum and energy along poloidal streamlines and magnetic
field lines (which are aligned with each other in ideal MHD). They are very useful
for diagnosing the mechanism for wind launching and acceleration
(e.g., \citealp{PelletierPudritz92,Spruit96}). In the shearing-box framework, the
counterparts of these conservation laws are derived by \citet{Lesur_etal12}. It was
found that poloidal gas streamlines do not necessarily follow exactly the poloidal
field lines, which introduces new terms in the conservation laws. Nevertheless, in
the absence of strong magnetic field, the deviations are small (as we checked in our
simulations) therefore, we just consider conventional terms in the treatment.

\begin{figure*}
    \centering
    \includegraphics[width=160mm]{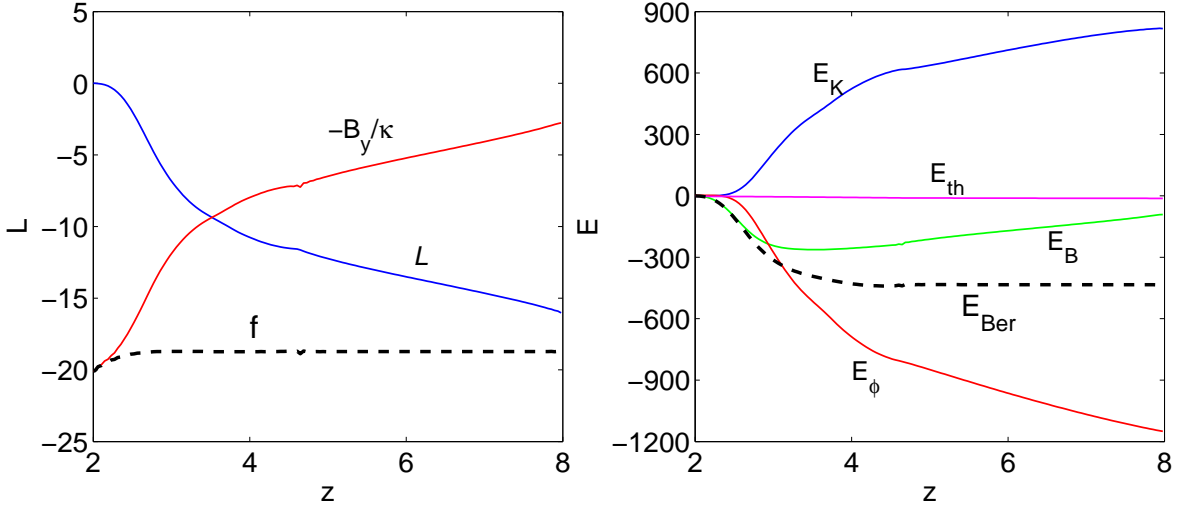}
  \caption{Angular momentum (left) and energy (right) conservation along a streamline
  in the disk surface from $z=2H$ to $8H$ for the laminar wind solution in our fiducial
  run OA-b5. For the energy plot, the various terms in Equation (\ref{eq:bernoulli}) are
  represented by: $E_{\rm Ber}$ (black dashed), $E_K$ (blue solid), $E_T$ (magenta
  solid), $E_B$ (green solid) and $\phi$ (red solid). For the angular momentum plot, the
  various terms in Equation (\ref{eq:angcons}) are represented by: $f$ (black dashed),
  ${\mc L}$ (blue solid), $-\bar{B}_y/\kappa$ (red solid).}\label{fig:conservation}
\end{figure*}

Starting from mass conservation, we have
\begin{equation}
\rho v_z={\rm const}\equiv \kappa B_z\label{eq:chi}
\end{equation}
at each side of the disk. In practice, because we constantly add mass to the simulation
box to maintain steady state, we find that $\kappa$ does
not become constant until beyond $z\sim\pm 3H$ for our run OA-b5.

Specific angular momentum is conserved along streamlines, as long as $\kappa$
is constant:
\begin{equation}
f\equiv{\mc L}-\frac{B_y}{\kappa}\ ,\label{eq:angcons}
\end{equation}
where ${\mc L}\equiv v_y+\Omega x/2$ is the fluid part of the specific angular
momentum. The partition between ${\mc L}$ and $B_y/\kappa$ describes
the angular momentum exchange between gas and magnetic field.

Energy conservation is expressed by the Bernoulli invariant in the ideal MHD
regime, and we expect it to strictly hold only beyond the FUV ionization front
(at $z\approx\pm4.7H$). It reads
\begin{equation}
\begin{split}\label{eq:bernoulli}
E_{\rm Ber}&=E_K+E_T+E_\phi+E_B\\
&=\frac{u^2}{2}+c_0^2\log(\rho)+\phi-\frac{B_yv_y^*}{\kappa}\ ,
\end{split}
\end{equation}
where $E_{\rm Ber}$ represents the specific energy along a streamline, with
the four terms denoting kinetic energy, enthalpy, potential energy and the work
done by the magnetic torque, respectively, and
\begin{equation}
v_y^*\equiv u_y-\kappa B_y/\rho\ .
\end{equation}
Note that the full velocity ${\mb u}$ (rather than ${\mb v}$) enters $E_K$, and
the potential energy for the shearing-box is
\begin{equation}
\phi=-\frac{3}{2}\Omega x^2+\frac{1}{2}\Omega z^2\ .
\end{equation}

To test these conservation laws, we integrate the velocity profiles shown in
Figure \ref{fig:FidOA-prof} to obtain the streamline. We start from $z=2H$
where $\kappa$ approaches the constant value and trace the streamlines to
the vertical boundary $z=8H$, setting $x=0$ at $z=2H$ (note that by
construction the conservation laws do not depend on the choice of the zero
point of $x$). This range covers the most interesting regions of the disk where
wind launching and acceleration take place.

In Figure \ref{fig:conservation} we show the vertical profiles of individual terms in
the specific angular momentum and energy for streamlines obtained from run
OA-b5. On the left panel, we see that the black dashed line flattens beyond
$z\approx2.5H$, indicating angular momentum conservation. This is consistent
with our expectations ($f$ is conserved once $\kappa$ approaches constant). It
is clear that the increase of gas angular momentum ($|{\mc L}|$) is compensated
by the magnetic torque. On the right panel, we see that the black dashed line
flattens beyond $z\gtrsim4H$, indicating energy conservation. This is again
consistent with our expectations ($E_{\rm Ber}$ is conserved once the gas obeys
ideal MHD, i.e., beyond the FUV ionization front). The rise of the blue curve in the
energy plot indicates flow acceleration, which is mostly compensated by the
reduction of potential energy (red). This is the basic picture of magnetocentrifugal
acceleration, which is not surprising since the bending angle of the field lines
relative to disk normal in the wind zone is about $45^\circ$, well above the
threshold of $30^\circ$. In this frame of reference, the magnetic torque plays a
minor role, although its effect and the centrifugal potential are interchangable by
shifting the zero point of $x$ \citep{BaiStone12a}. Thermal pressure gradient only
plays a diminishing role in the acceleration process.

\subsection[]{Angular Momentum Transport}\label{ssec:amt}

\begin{figure}
    \centering
    \includegraphics[width=90mm]{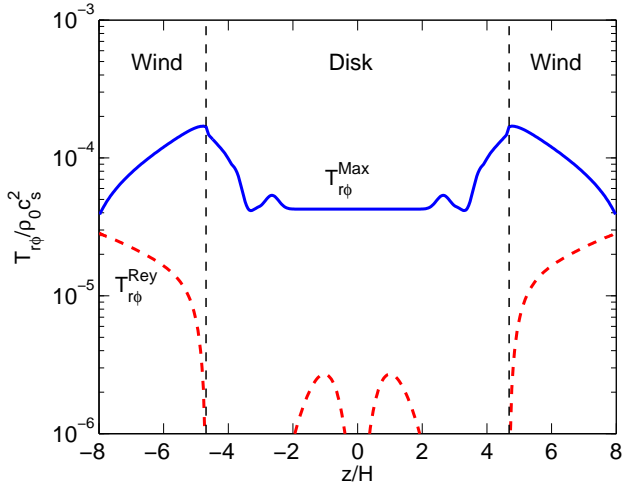}
  \caption{The vertical profiles of the $R\phi$ components of the Reynolds (red dashed)
  and Maxwell (blue solid) stresses in our fiducial run OA-b5. The disk is divided into the
  wind zone and disk zone, given by $v_y=0$.}\label{fig:radstress}
\end{figure}

Angular momentum transport can be achieved by radial redistribution, due to the
$R\phi$ component of the stress tensor as discussed in the beginning of Section
\ref{sec:fiducial}, as well as by vertical extraction, due to the $z\phi$ component of
the stress tensor
\begin{equation}
\begin{split}\label{eq:windT}
T_{z\phi}^{\rm Rey}&=(\rho v_yv_z)|_{\pm z_b}\ ,\\
T_{z\phi}^{\rm Max}&=(-B_yB_z)|_{\pm z_b}\ ,\\
\end{split}
\end{equation}
where subscripts $_{\pm z_b}$ corresponds to the location for the top and bottom
sides of the disk, which we choose to be at the base of the wind (the reason will
become clear shortly). The torque from the outflow can be obtained by simply
multiplying $T_{z\phi}$ by the radius $R$ (which is unspecified in shearing-box).

The total rate of angular momentum transport, assuming steady-state accretion in
the disk zone, can be obtained by generalizing equation (\ref{eq:mdot1}) to include
the wind torque, which gives
\begin{equation}
\begin{split}
\dot{M}=&\frac{2\pi}{\Omega}\int_{-z_b}^{z_b} dzT_{R\phi}
+\frac{4\pi}{\Omega}RT_{z\phi}\bigg|_{-z_b}^{z_b}\\
=&\frac{2\pi}{\Omega}\alpha c_s^2\Sigma
+\frac{8\pi}{\Omega}R|T_{z\phi}|_{z_b}\ ,\label{eq:mdot}
\end{split}
\end{equation}
where we have assumed $T_{z\phi}|_{z_b}=-T_{z\phi}|_{-z_b}$ from symmetry
considerations, which will further be discussed in the next subsection. We see
that if $T_{R\phi}$ and $T_{z\phi}$ are comparable with each other, than vertical
transport would be more efficient than radial transport by a factor of about $R/H$.
Since we are considering thin accretion disks (where shearing-box approximation
applies), it is in general much easier for disk wind to transport angular
momentum compared with turbulent transport.

We start by examining the radial transport of angular momentum. Although radial
transport is usually a result of turbulence which gives correlated fluctuations of
velocity and magnetic field in the radial and azimuthal direction, ordered
velocity/field structure also produce radial transport. In Figure \ref{fig:radstress},
we show the vertical profiles of the $R\phi$ components of the Reynolds and
Maxwell stresses. Clearly, even there is no turbulence, a non-zero Maxwell stress
given by ordered magnetic fields dominates the radial transport in the disk zone.
This corresponds to the undead zone scenario discussed in \citet{TurnerSano08}.
We also see that contribution from the Reynolds stress in the disk zone is
completely negligible. Integrating the Maxwell stress across the disk zone, we find
$\alpha\approx\alpha_{\rm Max}\approx2.3\times10^{-4}$.

Meanwhile, Figure \ref{fig:radstress} best demonstrates the advantage for dividing
the profile of our the laminar wind solution into the disk zone and the wind zone:
the Reynolds stress is by definition zero at the base of the wind, and increases
rapidly in the wind zone as a result of radially-outward and super-Keplerian motion
consequent of the magnetocentrifugal acceleration; the Maxwell stress has a
prominent peak at the base of the wind, and decreases in the wind zone, as
magnetic energy is converted to the kinetic energy of the outflowing gas. 

For the vertical transport, we measure the $z\phi$ component of the stress tensor
$T_{z\phi}$ at the base of the wind. Again, our choice of the wind base shows its
advantages: by definition, Reynolds stress is zero, and one only needs to consider
the Maxwell stress, which we find is
$T_{z\phi}^{\rm Max}|_{z_b}\approx1.07\times10^{-4}\rho_0c_s^2$. Hereafter, we will
refer to the $z\phi$ component of the Maxwell stress at the base of the wind as
``wind stress". We note that since $T_{z\phi}^{\rm Max}=-B_zB_\phi$ where $B_z$
is constant, from Figure \ref{fig:FidOA-prof} we see that $T_{z\phi}^{\rm Max}$ varies
smoothly within the disk and should peak between $z=\pm2.5H$. Although wind stress
eventually drives accretion, a small fraction of the gas flow undergoes ``decretion" to
bend magnetic field lines (between the launching point and the wind base, see Figure
\ref{fig:fieldline}). This outwardly directed gas flow is properly accounted for by
measuring the wind stress at $z=\pm z_b$ rather than at $z\sim\pm2.5H$.

Applying the MMSN disk model into Equation (\ref{eq:mdot}), we can estimate the
accretion rate driven by the combined effect of radial and vertical transports
\begin{equation}
\dot{M}_{-8}\approx0.82\bigg(\frac{\alpha}{10^{-3}}\bigg)R_{\rm AU}^{-1/2}
+4.1\bigg(\frac{|T_{z\phi}|_{z_b}}{10^{-4}\rho_0c_s^2}\bigg)R_{\rm AU}^{-3/4}\ ,\label{eq:accrate}
\end{equation}
where $\dot{M}_{-8}$ is the accretion rate measured in $10^{-8}M_{\bigodot}$
yr$^{-1}$. We see that radial transport by the Maxwell stress (from ordered
magnetic field) in the disk zone can drive accretion for about
$2\times10^{-9}M_{\bigodot}$ yr$^{-1}$. This number is already significantly
larger than the case for run OA-nobz with zero net vertical magnetic flux, yet
still too small to account for the typical accretion rate for PPDs. The accretion
driven by wind transport, on the other hand, reaches
$4.3\times10^{-8}M_{\bigodot}$ yr$^{-1}$, which is sufficient to account for
the observed accretion rate for most PPDs.

We see that our laminar wind solution simultaneously solves two problems
facing the conventional MRI scenario and the conventional wind scenario
described in Section \ref{sec:intro}. First, it efficiently drives disk accretion
that matches the rates inferred from observations, which can hardly be
achieved in the MRI scenario. Second, it launches the magnetocentrifugal
wind with only a very weak net vertical magnetic field, in contrast with the
conventional understanding that requires equipartition field at disk midplane.
Therefore, the rate of wind transport in our scenario is only moderate and
does not result in rapid depletion of the disk.

\subsection{Symmetry and Strong Current Layer}\label{ssec:symmetry}

\begin{figure}
    \centering
    \includegraphics[width=90mm]{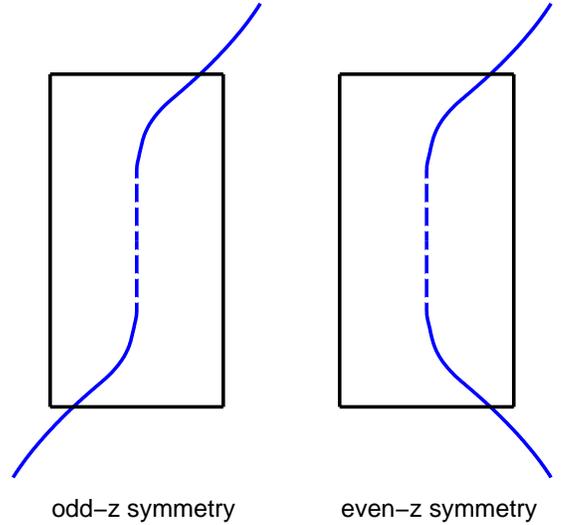}
  \caption{Cartoon illustration of the geometry of wind magnetic field lines in
  shearing-box simulations. The shearing-box approximation ignores all radial
  gradients (except the shear), hence does not contain information about the
  location of the central object. The natural wind geometry launched from
  shearing-box has the ``odd-$z$" symmetry where the field lines on the top and
  bottom of the box bend toward opposite directions (left). In a physical situation,
  where the location of the central object is fixed, the field lines on the top and bottom
  sides of the box should bend to the same direction away from the central object.}\label{fig:symmetry}
\end{figure}

The previous subsection demonstrates the success of the magnetocentrifugal wind
scenario as the the driving force of disk accretion. However, one problem was ignored
in the previous discussion. The laminar wind solution we obtained obeys the ``odd-$z$"
symmetry:
\begin{equation}
\begin{split}\label{eq:oddz}
B_x(z)=B_x(-z)\ ,\quad v_x(z)=-v_x(-z)\ ,\\
B_y(z)=B_y(-z)\ ,\quad v_y(z)=-v_y(-z)\ ,\\
B_z(z)=B_z(-z)\ ,\quad v_z(z)=-v_z(-z)\ .\\
\end{split}
\end{equation}
As a result, the wind field lines on the top and bottom sides of our simulation box
bend to opposite radial and toroidal directions, as illustrated on the left of the
Cartoon picture in Figure \ref{fig:symmetry}. However, physically, one expects the
field lines on the top and bottom sides of the box to bend toward the same direction,
pointing away from the central star, as illustrated on the right of Figure
\ref{fig:symmetry}. A particularly disturbing fact is that in the case with odd-$z$
symmetry, the ``wind" does not transport angular momentum, since $T_{z\phi}$ at
the top and bottom sides of the disk have the same sign and value, and cancel out.
The wind angular momentum transport mechanism works only for a physical wind
geometry.

We note that all radial gradients except the azimuthal shear are neglected in the
shearing-box approximation, hence one can not tell whether the location of the
central star is in the ``inner" or ``outer" (radial) side of the box. In other words, the
two radial sides of the box are symmetric. In our simulations, we find that the
direction that the wind field lines bend as we start from the initial configuration is
totally random, and the top and bottom sides of the disk evolve
independently\footnote{We have performed more than one copies of each 3D and
the quasi-1D fiducial simulations that verifies the randomness.}.
This suggests that the chances to find a solution with the unphysical odd-$z$
symmetry and with a physical geometry are equal. In fact, our 1D run S-OA-b5
results in a solution with the physical geometry.

\begin{figure*}
    \centering
    \includegraphics[width=160mm]{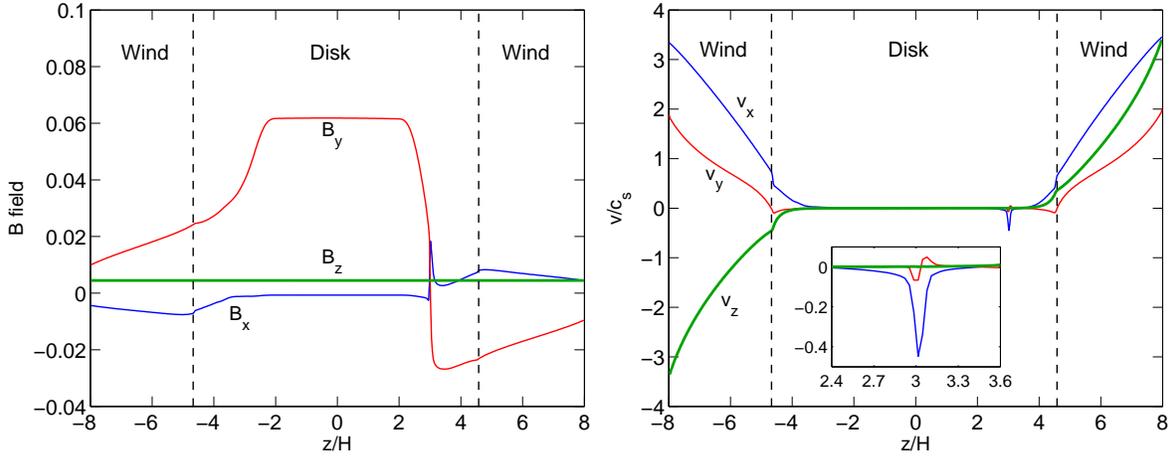}
  \caption{The vertical profiles of the magnetic fields (left) and velocities (right)
  in the laminar solution that has the physical wind geometry (run S-OA-b5, see
  Section \ref{ssec:symmetry}). The inset on the right panel shows the zoomed-in
  view of the velocity profiles at the strong current layer. This is essentially how
  accretion proceeds in our wind-driven scenario.}\label{fig:phyprof}
\end{figure*}

In Figure \ref{fig:phyprof}, we show the profiles of the magnetic field and
velocity field from our run S-OA-b5-F3 that have the physical wind geometry. We
see that this physical solution is NOT a solution under the ``even-$z$" symmetry:
\begin{equation}
\begin{split}\label{eq:evenz}
B_x(z)=-B_x(-z)\ ,\quad v_x(z)=v_x(-z)\ ,\\
B_y(z)=-B_y(-z)\ ,\quad v_y(z)=v_y(-z)\ ,\\
B_z(z)=B_z(-z)\ ,\quad v_z(z)=-v_z(-z)\ .\\
\end{split}
\end{equation}
which is almost exclusively used for constructing semi-analytical local wind solutions
\citep{WardleKoenigl93,OgilvieLivio01,Konigl_etal10,Salmeron_etal11}. Very
interestingly, the physical wind solution follows exactly the odd-$z$ symmetry solution
we obtained before, as can also be seen from Table \ref{tab:windresults} that all
physical quantities measured in this run agree with those in our 3D fiducial run OA-b5
within $5\%$. In particular, the wind stress $T_{z\phi}^{\rm Max}$ is almost exactly the
same. The only difference is that the horizontal field lines (and horizontal velocities)
are flipped at one side of the box to achieve the physical wind geometry. This flipping
is mediated by a sharp transition of the horizontal fields, which exhibits as a strong
current layer. The strong current layer is not located at the midplane, but is located at
about $z=+3H$ (or at $z=-3H$, and the selection is random). Therefore, the physical
solution does not have any symmetry about the disk midplane.

The location of the strong current layer roughly corresponds to where both the
Ohmic and AD Elsasser numbers $\Lambda$ and $Am$ become greater than $1$,
as seen from the bottom left panel of Figure \ref{fig:FidOA-prof}. The reason for
such large offset from the disk midplane is that magnetic diffusion near the disk
midplane is so strong that the gas and magnetic field are essentially decoupled.
Correspondingly, the magnetic field lines must be straight and current is excluded,
and the strong current layer can exist only when magnetic diffusion becomes weaker.

The right panel of Figure \ref{fig:phyprof} demonstrates the velocity profile
of the physical wind solution. The strong current layer exhibits as a seemingly
small feature at about $z=+3H$ in this plot, which is in linear scale. As we zoom
into this region shown in the inset, we find that the gas has a large inflow velocity
of about $0.3c_s$ at the location of the strong current layer. The large inflow is
directly the consequence of the wind stress: the Maxwell stress exerted at the base
of the wind is released at the strong current layer, driving a large inflow which
is essentially how accretion proceeds. More specifically, it is the balance between
the Lorentz force and Coriolis force that leads to the large inflow and accretion in
the strong current layer (see equation (\ref{eq:balance})).

To further demonstrate the effectiveness of accretion in the current layer, we note
that for a MMSN disk, in order to have accretion to approach $10^{-8}M_{\bigodot}$
yr$^{-1}$, a {\it bulk} radial drift velocity of about $10^{-4}c_s$ is required
\begin{equation}
\dot{M}_{-8}\approx2.5\bigg(\frac{v_r}{10^{-4}c_s}\bigg)R_{\rm AU}^{-3/4}\ .\label{eq:vr}
\end{equation}
In the physical wind solution, we find that only a tiny fraction (about
$\sim5\times10^{-4}$) of disk mass is contained in the strong current layer, but it
drifts at very large velocities ($\sim0.3c_s$). The combined effect is an efficient
accretion that easily accounts for the typical accretion rates observed in PPDs.
More specifically, by integrating the radial mass flux ($\int\rho v_x dz$) across
the disk zone, we find the mean radial velocity of the gas to be
$-1.69\times10^{-4}c_s$. Plugging to Equation (\ref{eq:vr}), this would lead to an
accretion rate of $4.2\times10^{-8}M_{\bigodot}$ yr$^{-1}$, which agrees very well
with the estimate in Section \ref{ssec:amt}.

Finally, we note that the strong current layer is not a thin current sheet, since at its
location there is still substantial magnetic diffusion (mainly AD). It has a thickness of
about $0.25H$, which is well resolved in about $6$ cells in our quasi-1D simulations.
We have found that the strong current layer is stable in our quasi-1D fiducial run
S-OA-b5. The stability of the strong current layer in 3D requires justification in the
more geometrically appropriate global simulations\footnote{We have repeated our
3D fiducial run OA-b5 with a different random seed which evolves to a physical wind
solution with a strong current layer. We find that the strong current layer is maintained
for about 100 orbits but eventually escapes the simulation box and the odd-$z$ symmetry
is recovered. This is likely to be due to the intrinsic limitations of the shearing-box,
where curvature terms are neglected and hence favors the odd-$z$ symmetry solution.
Moreover, as the fast magnetosonic point is beyond our simulation box, the stability
of the strong current layer can also be affected from outside of the simulation domain.}.

\subsubsection[]{Wind Solution from Even-$z$ Symmetry}\label{sssec:evenz}

\begin{figure*}
    \centering
    \includegraphics[width=160mm]{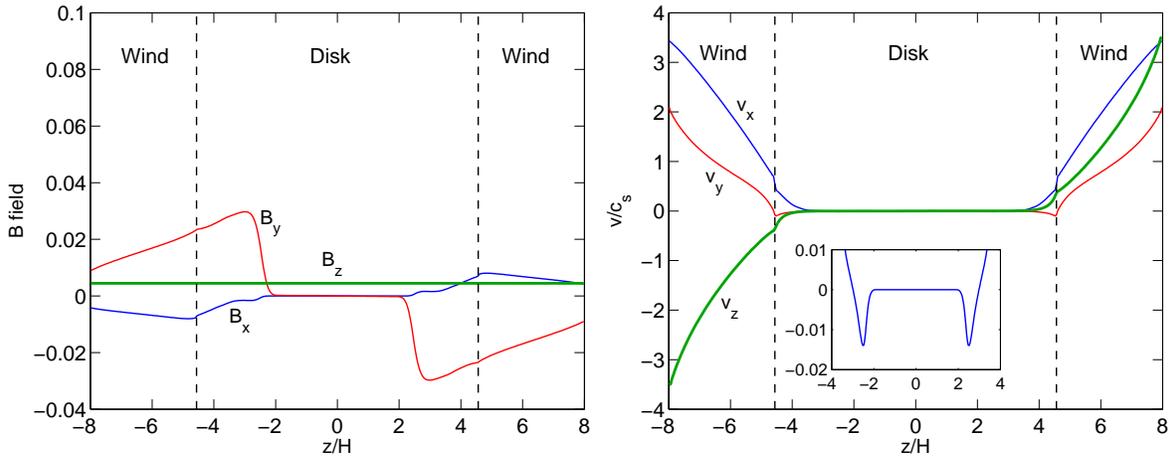}
  \caption{Same as Figure \ref{fig:phyprof}, but for run E-OA-b5 where an even-$z$
  symmetry (\ref{eq:evenz}) is enforced (see Section \ref{sssec:evenz}). The inset on
  the right panel shows the zoomed-in view of the radial velocity profile in the disk
  zone.}\label{fig:evenzprof}
\end{figure*}

We have discussed that the odd-$z$ symmetry is the most natural symmetry in
shearing-box but it is unphysical for a disk wind. A physical wind solution can be
achieved by flipping the horizontal field and stream lines at one side of the disk,
which yields a solution that contains a strong current layer offset from the disk
midplane. Because the two solutions have exactly the same properties except
the flip, it is natural to classify them as the same solution. In this subsection, we
further consider a solution that obeys even-$z$ symmetry.

We design a new 3D simulation E-OA-b5 which is exactly the same as our fiducial
run OA-b5 except that we use only half of the shearing-box, with the lower
boundary located at the disk midplane. At the lower boundary we use the standard
reflection/conduction boundary condition so that even-$z$ symmetry is enforced.
We find that after a brief period of the MRI, the system also settles into a pure
laminar state, with velocity and magnetic field profiles shown in Figure
\ref{fig:evenzprof} (where we have recovered the other side of the disk based on
even-$z$ symmetry). Under even-$z$ symmetry, the horizontal magnetic field at the
midplane is enforced to be zero. Due to the large resistivity (which can not support
a significant current), the field is kept very close to zero within $z\sim\pm2H$ where
both Ohmic and AD Elsasser numbers are much less than $1$. Beyond $z=\pm2H$,
the horizontal field increases, and then beyond $z\sim\pm3H$, the field configuration
recovers to almost exactly the same solution we obtained under odd-$z$ symmetry.

We note that the field configuration similar to our even-$z$ symmetry solution has
previously been reported by \citet{Li96} and \citet{Wardle97}, and it was discussed
that in the presence of low conductivity in the midplane, wind solution can be
constructed with midplane magnetic field pressure much smaller than the gas thermal
pressure. Our solution confirms their findings and demonstrates it robustness and
uniqueness since it is constructed in a self-consistent and evolutionary manner.

The velocity profiles from our even-$z$ symmetry solution are exactly the same as
the odd-$z$ symmetry solution in the wind zone (except the flip). In the disk zone,
we see that the radial velocity has two dips located at $z\sim\pm3H$. The dips are
much shallower (only about $0.01c_s$) than the dip due to the strong current layer
in our run S-OA-b5, but they are much wider. Integrating the $\rho v_x$ over height,
we obtain the radial mass flux to be $-1.67\times10^{-4}c_s$, which is almost exactly
the same as in run S-OA-b5. This is not surprising since the properties of the wind
in the two solutions are exactly the same.

In sum, we find that the suppression of the MRI and wind launching is inevitable in
the inner region of PPDs around 1AU, and the property of the solution in the wind
zone is independent of the large-scale field geometry (or symmetry).

\subsection[]{Outflow}\label{ssec:outflow}

Strong outflows are launched in our fiducial simulations, and we have measured
that the rate of mass outflow from each side of the box is about
$1.4\times10^{-5}\rho_0c_s$ from the quasi-1D run and $1.5\times10^{-5}\rho_0c_s$
from the 3D run. In Table \ref{tab:windresults}, we also list total mass loss rate,
\begin{equation}
\dot{M}_w=|\rho v_z|_{\rm bot}+|\rho v_z|_{\rm top}
\end{equation}
as the sum of the mass loss rate from the two vertical boundaries for our fiducial
runs. Although mass loss is not significant for the duration of our
simulations, the value is relatively high that without replenishing to the disk, the
mass loss timescale is only $8\times10^4\Omega^{-1}$, which amounts to only
about $10^4$ years. Moreover, the measured wind mass loss rate is in fact
comparable to the mass accretion rate driven by the wind itself, which is
inconsistent with the observationally inferred ratio that the mass loss rate is only
about $10\%$ of the accretion rate \citep{Cabrit_etal90,Hartigan_etal95}.

We note that, however, the magnetocentrifugal wind is intrinsically a global
phenomenon. A global wind solution is not fully determined until all critical
points (or surfaces), namely, the slow magnetosonic, Alfv\'en and fast
magnetosonic points are passed, where the fast magnetosonic point is far
beyond the extent of our simulation box. This leads to one extra degree of
freedom in the wind solution. Moreover, the global structure of the
wind, and the location of the critical points, are set by the interplay between
the microphysics in the disk, and the large-scale field structure (i.e., the
magnetic flux distribution over the entire disk), where the latter is not reflected
in local shearing-box simulations. Therefore, while useful for studying
wind-launching, the shearing-box approximation has its limitations and the
measured rate of outflow from our simulations should only be taken as reference
and it may significantly overestimate the mass outflow rate in reality.

To further clarify the discussion above, we perform three additional quasi-1D
simulations, changing the box size to $12H$, $20H$
and $24H$ respectively, and the three runs are labeled as S-OA-b5-12H,
S-OA-b5-20H and S-OA-b5-24H. We measure a series of quantities discussed
before and also list the results in Table \ref{tab:windresults}. We
find that as we increase the height of the simulation box, the location of the Alfv\'en
point moves to higher altitudes. In the mean time, the mass loss rate decreases.
In fact, these two quantities are connected, and in shearing-box, the definition of
the Alfv\'en point indicates \citep{BaiStone12a}
\begin{equation}
\dot{M}_w=2\sqrt{\rho}\bigg|_A\cdot|B_z|\ ,
\end{equation}
where $|_A$ denotes the value at the Alfv\'en point. Therefore, higher location of the
Alfv\'en point indicates lower density, hence smaller mass loss rate.

The trend of deceasing mass outflow rate with vertical box size reflects one limitation
to shearing-box: the gravitational potential increases quadratically with vertical height,
which increasingly hinders the gas from escaping. The outflow in our simulations is
possible because gravitational potential is cut off at the vertical boundaries. This
limitation of shearing-box has been noticed and discussed in a number of recent works
on ideal MHD simulations of the MRI \citep{Fromang_etal12,Lesur_etal12}. Because
the depth of the shearing-box potential is about the same as the real depth of the potential
at $z\sim\pm R$,
\citet{BaiStone12a} suggest that the mass loss rate can be roughly estimated by performing
shearing-box simulations with a box height of $L_z\sim2R$, or to study the height
dependence of $\dot{M}_w$ and extrapolate the result to $L_z\sim2R$. In MMSN, we
have $H/R\sim0.03$ at 1 AU (note that $H/R$ is not specified in the shearing-sheet
approximation). Such a thin disk would require a shearing-box height of $L_z\sim65H$,
which is numerically unpractical. Based on our three quasi-1D runs with $L_z=12H$ to
$24H$, the trend can be roughly described by $\dot{M}_w\sim1/L_z$.
Extrapolating this trend, we obtain a rough estimate of mass loss rate to be
$\dot{M}_w\sim6\times10^{-6}$ in code unit. This level of $\dot{M}_w$ brings the mass
loss time scale to about $10^5$ years, which is about an order of magnitude longer than
our original naive estimate. Although still somewhat short, the mass reservoir in the outer
disk (which contains much more mass than that in the inner disk) may be sufficient to feed
the inner disk to compensate for the mass loss.


It is also interesting to note that although the mass outflow rate is largely affected by the
vertical box size, the angular momentum transport depends on the vertical box size very
weakly. For box sizes of $12H$, $16H$, $20H$ and $24H$, the values of
$T_{z\phi}^{\rm Max}$ at the base of wind are found to be $0.97\times10^{-4}$,
$1.06\times10^{-4}$, $1.13\times10^{-4}$ and $1.18\times10^{-4}$ respectively, hence
increasing box size leads to more or less the same (but slightly stronger) wind-driven
accretion.

In sum, the trend that the mass outflow rate decreases with box size and mass accretion
rate increases with box size implies that in a real system, the ratio of mass accretion rate
to mass loss rate would be a factor of several larger than that obtained in our simulations,
and would be more consistent with observations.


\subsection[]{Energetics}\label{ssec:energetics}

Although our simulations assume an isothermal equation of state where total energy
is not conserved, it is important to verify that our simulation results are energetically
feasible and to examine the energy balance for real situations.

In shearing-box simulations, the energy is injected due to the work done by the
shearing-box boundaries, the rate of which is given by \citep{BaiStone12a}
\begin{equation}
P_{\rm sh}=\frac{1}{L_xL_y}\int\frac{3}{2}\langle \rho v_xv_y
-B_xB_y\rangle\Omega L_xdydz=\frac{3}{2}\Sigma c_s^2\Omega\alpha\ ,
\end{equation}
where the integral over $z$ is performed within $z=\pm z_b$ (the disk zone).
Energy is lost through the open vertical boundaries, given by
\begin{equation}
\dot{E}=\dot{E}_K+\dot{E}_P=\bigg[\rho\bigg(\frac{v^2}{2}+c_s^2\bigg){\mb v}
-({\mb v}\times{\mb B})\times{\mb B}\bigg]\cdot{\mb n}\bigg|_{z=\pm z_b}\ ,
\end{equation}
where ${\mb n}$ is the unit vector pointing away from the disk in the vertical direction,
and we sum over contributions from the top and bottom of the wind base.
The first terms in the bracket represent the kinetic energy loss and the $PdV$ work
done by the mass outflow, and the second term represents energy loss from the
Poynting flux. We have ignored the energy loss term associated with internal energy
loss due to the mass outflow, since we keep feeding mass to the system which
balances this term exactly.

In Table \ref{tab:windresults} we also list the values of $P_{\rm sh}$, $\dot{E}_K$
and $\dot{E}_P$, normalized in natural units. We see that the sum of $\dot{E}_K$
and $\dot{E}_P$ comprises of about $60\%$ of the total work done by the shear
$P_{\rm sh}$, hence energy conservation is not violated. The extra energy is likely
to be radiated away in real systems.

\subsection[]{Radial Drift of Magnetic Flux}

\begin{figure}
    \centering
    \includegraphics[width=80mm]{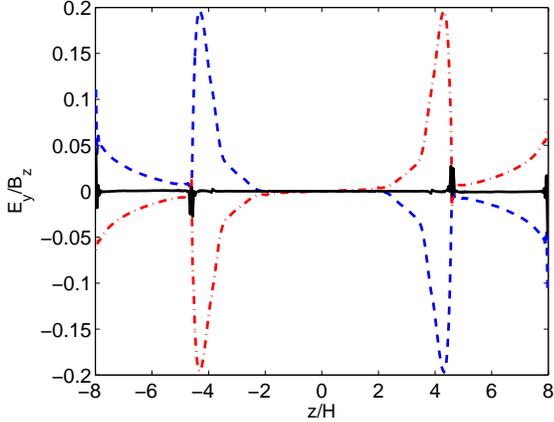}
  \caption{The toroidal EMF profile in our fiducial run OA-b5. Blue dashed:
  inductive EMF (${\mc E}^I_y$); red dash-dotted: AD EMF (${\mc E}^A_y$);
  black solid: the total EMF (${\mc E}_y$). The Ohmic EMF is very close to
  $0$ and is not plotted. The EMFs are normalized to $B_z$ and are in natural
  unit of the sound speed $c_s$.}\label{fig:emf}
\end{figure}

Globally, poloidal magnetic flux drifts radially at velocity $v_B$ in the presence of
a net toroidal electromotive force (EMF, ${\mc E}$)
\begin{equation}
v_B=\frac{-{\mc E}_y}{B_z}\ .
\end{equation}
Negative $v_B$ would lead to accumulation of magnetic flux in the inner disk
while positive $v_B$ would make the disk lose magnetic flux to large scales.
Moreover, ${\mc E}_y$ should be constant with height so that magnetic flux
drifts uniformly across the disk.
In global models, the radial drift velocity $v_B$ is determined by the radial
distribution of magnetic flux \citep{Teitler11}. In local models, it is often chosen
as a free parameter in local shearing-box models of disk winds
\citep{WardleKoenigl93} due to the extra degree of freedom discussed in
Section \ref{ssec:outflow}. Our simulations in principle have one extra degree of
freedom because the fast magnetosonic point is not contained in the simulation
box. We expect that the value of $v_B$ in our simulations has direct
correspondence with this extra degree of freedom.

The toroidal component of the EMF in our simulations is the sum of the inductive
EMF (${\mc E}^I$), Ohmic EMF (${\mc E}^O$) and AD EMF (${\mc E}^A$),
which reads
\begin{equation}
\begin{split}
{\mc E}_y&={\mc E}_y^I+{\mc E}_y^O+{\mc E}_y^A\\
&=(v_xB_z-v_zB_x)+\eta_OJ_y+\eta_A({\mb J}_{\perp})|_y\ .\\
\end{split}
\end{equation}
In Figure \ref{fig:emf} we show the radial profiles of the three toroidal EMFs
and their sum, all normalized to $B_z$. We see that in most regions of the
disk, the inductive EMF is essentially exactly balanced by the AD EMF. We
did not plot the Ohmic EMF because it is always negligibly small due to the
very small current near the midplane and the very small resistivity in the
upper layer. The sum of all EMFs, is consistent with zero at all locations of
the disk. The little sharp peaks of the total EMF at around
$z\approx\pm4.7H$ are purely numerical features due to the sharp variations
of the AD coefficient and current density at the FUV ionization front. 

Therefore, the wind solution we obtained from shearing-box simulations
is a zero radial drift solution $v_B=0$, which corresponds to a stationary
magnetic configuration in the global picture. A direct consequence of
$\epsilon_B=0$ is that poloidal field lines and poloidal streamlines follow
each other in the absence of non-ideal MHD terms, as clearly shown in
Figure \ref{fig:fieldline}. The fact that $v_B=0$ in our simulations is likely to
be a result of our vertical boundary condition, where there is no vertical
gradient of magnetic fields. We note that in the ideal MHD shearing-box
simulations of wind launching process by \citet{Lesur_etal12}, significant
deviation between poloidal field lines and streamlines are found (see their
Figure 3). This is likely to be a result of a different vertical boundary
condition: poloidal field is forced to be vertical at ghost zones, resulting in
a current sheet at the vertical boundary.

\section[]{Parameter Study}\label{sec:parameter}

To assess the importance of various physical effects on the properties of the
laminar wind solution, we conduct a parameter study by surveying four
parameters and compare the results from our fiducial model. All simulations in
this parameter study are quasi-1D. First, we vary the vertical net magnetic flux,
and consider $\beta_0=10^3$, $10^4$ and $10^6$. These runs are labeled as
S-OA-b$n$, where b$n$ denotes $\beta=10^n$. Second, we vary the
penetration depth of the FUV ionization, and consider
$\Sigma_{\rm FUV}=0.01$ g cm$^{-2}$ and $0.1$ g cm$^{-2}$, and the run
labels are attached with ``F1" and ``F10" respectively. Third, we consider the
effect of grain abundance, and consider the case with grain-free chemistry
(whose run label is attached with ``nogr") and the case with grain abundance
$\epsilon_{\rm gr}=10^{-3}$ (run label attached with ``gr01", indicating depletion
factor of 0.1), 10 times the value in our fiducial run. Finally, we vary the surface
density of our disk model, and consider $\Sigma=3\Sigma_{\rm MMSN}$ and
$\Sigma=0.3\Sigma_{\rm MMSN}$. In the mean time, we maintain the same
field strength in {\it physical} unit, thus $\beta_0$ in these two runs are set to
$3\times10^5$ and $3\times10^4$ respectively. These two runs are attached
with ``M3" and ``M03", respectively.
A list of all these runs are given in Table \ref{tab:paramruns}.

\begin{table}
\caption{Summary of All Simulations in the Parameter Study.}\label{tab:paramruns}
\begin{center}
\begin{tabular}{ccccc}\hline\hline
 Run & $\Sigma/\Sigma_{\rm MMSN}$ & $\beta_0$ &
 $\Sigma_{\rm FUV}$ & $\epsilon_{\rm gr}$ \\\hline
S-OA-b3 & 1 & $10^3$ & 0.03 & $10^{-4}$ \\
S-OA-b4 & 1 & $10^4$ & 0.03 & $10^{-4}$ \\
S-OA-b6 & 1 & $10^6$ & 0.03 & $10^{-4}$ \\
S-OA-F1 & 1 & $10^5$ & 0.01 & $10^{-4}$ \\
S-OA-F10 & 1 & $10^5$ & 0.1 & $10^{-4}$ \\
S-OA-nogr & 1 & $10^5$ & 0.03 & 0 \\
S-OA-gr01 & 1 & $10^5$ & 0.03 & $10^{-3}$ \\
S-OA-M3 & 3 & $3\times10^5$ & 0.1 & $10^{-4}$ \\
S-OA-M03 & 0.3 & $3\times10^4$ & 0.1 & $10^{-4}$ \\
\hline\hline
\end{tabular}
\end{center}
\end{table}

\begin{table*}
\caption{Wind Properties from All Simulations in the Parameter Study.}\label{tab:windresults2}
\begin{center}
\begin{tabular}{ccccccccc}\hline\hline
 Run & $\alpha_{\rm Max}$ & $T_{z\phi}^{\rm Max}$ & $\dot{M}_w$
 & $z_b$ & $z_{\rm A}$ & $P_{\rm sh}$ & $\dot{E}_P$ & $\dot{E}_K$\\\hline

S-OA-b3 & $1.19\times10^{-2}$  &  $3.37\times10^{-3}$ & $4.76\times10^{-4}$
& 3.95 & $>$8.00 & $6.93\times10^{-2}$ & $4.29\times10^{-3}$ & $3.52\times10^{-2}$ \\

S-OA-b4 & $1.27\times10^{-3}$  & $5.90\times10^{-4}$ & $8.70\times10^{-5}$
& 3.92 & 7.47 & $9.31\times10^{-3}$ & $2.42\times10^{-3}$ & $3.34\times10^{-3}$ \\

S-OA-b5 & $2.42\times10^{-4}$ & $1.06\times10^{-4}$ & $2.86\times10^{-5}$
& 4.63 & 6.22 & $2.02\times10^{-3}$ & $8.05\times10^{-4}$ & $4.03\times10^{-4}$ \\

S-OA-b6 & $3.01\times10^{-5}$ & $2.86\times10^{-5}$ & $1.00\times10^{-5}$
& 4.52 & 5.52 & $4.62\times10^{-4}$ & $2.14\times10^{-4}$ & $4.17\times10^{-5}$ \\

S-OA-F1 & $1.91\times10^{-4}$ & $9.16\times10^{-5}$ & $1.62\times10^{-5}$
& 4.17 & 7.11 & $1.61\times10^{-3}$ & $6.14\times10^{-4}$ & $2.81\times10^{-4}$ \\

S-OA-F10 & $1.62\times10^{-4}$ & $1.66\times10^{-4}$ & $5.26\times10^{-5}$
& 4.20 & 5.52 & $2.60\times10^{-3}$ & $1.09\times10^{-3}$ & $5.39\times10^{-4}$ \\

S-OA-gr01 & $1.32\times10^{-4}$ & $1.04\times10^{-4}$ & $2.69\times10^{-5}$
& 4.55 & 6.30 & $1.63\times10^{-3}$ & $7.83\times10^{-4}$ & $3.92\times10^{-4}$ \\

S-OA-nogr & $4.96\times10^{-4}$ & $1.16\times10^{-4}$ & $3.84\times10^{-5}$
& 4.92 & 5.86 & $2.87\times10^{-3}$ & $9.03\times10^{-4}$ & $4.62\times10^{-4}$\\

S-OA-M3 & $8.04\times10^{-5}$ & $3.64\times10^{-5}$ & $1.06\times10^{-5}$
& 4.89 & 6.23 & $6.50\times10^{-4}$ & $2.67\times10^{-4}$ & $1.38\times10^{-4}$\\

S-OA-M03 & $7.34\times10^{-4}$ & $3.51\times10^{-4}$ & $9.17\times10^{-5}$
& 4.38 & 6.09 & $6.60\times10^{-3}$ & $2.79\times10^{-3}$ & $1.36\times10^{-3}$\\
\hline\hline
\end{tabular}
\end{center}
Note: we have duplicated run S-OA-b5 as a standard reference.
\end{table*}

\begin{figure*}
    \centering
    \includegraphics[width=160mm]{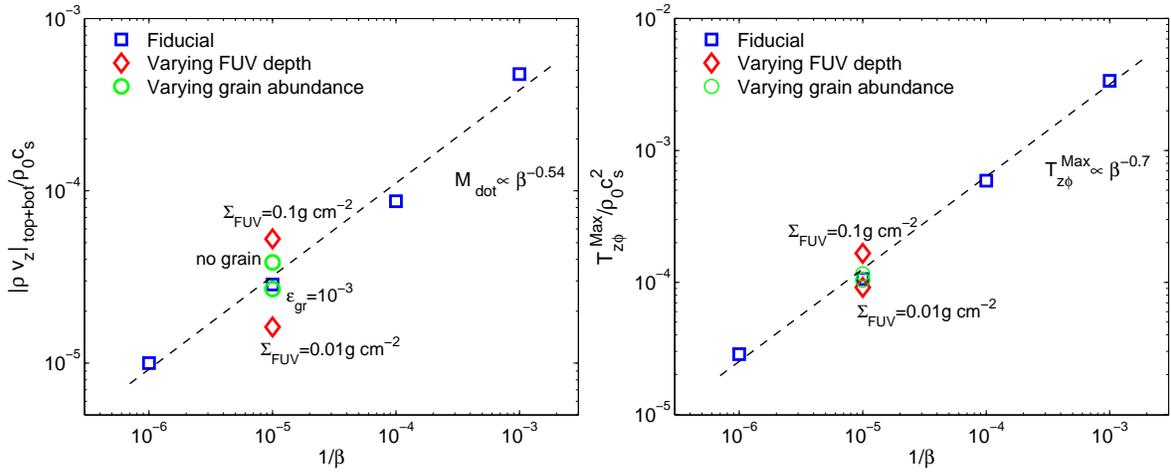}
  \caption{The rate of mass outflow (left) and the $z\phi$ component of the Maxwell stress
  at the base of the wind (right) as a function of vertical net field strength, indicated as
  $1/\beta_0$ for all our simulation runs. Particularly, blue squares represent simulations
  with varying vertical field strength, red diamonds represent simulations with varying FUV
  penetration depth, and green circles represent simulations with varying grain
  abundance.}\label{fig:parameter}
\end{figure*}

All the additional quasi-1D simulations are run to $t=1200\Omega^{-1}$. We extract
the vertical profiles by time averaging the second half of the runs and perform the
same analysis as we did for the fiducial run. In particular, we identify the locations of
the wind base and the Alfv\'en point, the $\alpha$ parameter of the Maxwell stress
in the disk zone, the $z\phi$ component of the Maxwell stress at wind base,
and so on. The results of these measurements are listed in Table \ref{tab:windresults2}.
Before discussing these physical effects in details, we plot in Figure \ref{fig:parameter}
the rate of mass outflow and the wind stress as a function of net vertical field strength.

\subsection{Effect of Net Vertical Magnetic Flux}

The vertical net magnetic field is a crucial ingredient to launch a magnetic
outflow. It allows the field lines to be connected to infinity, paving way for
the outflow to escape from the disk. We see from Figure \ref{fig:parameter}
that the rate of mass outflow $\dot{M}_w$ increases rapidly with increasing net
vertical field. A fit to the scattered plot gives $\dot{M}_w\propto\beta_0^{-0.54}$.
In other words, we roughly have $\dot{M}_w\propto B_z$. This result is
consistent with the ideal MHD shearing-box simulations by \citet{SuzukiInutsuka09},
as well as more recently by \citet{BaiStone12a}. Although the wind launching is
associated with vigorous MRI turbulence in ideal MHD, the scaling of the mass
loss rate with vertical background magnetic field is very similar.

On the right panel of Figure \ref{fig:parameter}, we see that the rate of
angular momentum transport, characterized by the wind stress, increases even
more rapidly with background vertical field strength. A simple fit returns
$T_{z\phi}^{\rm Max}\propto\beta_0^{-0.7}$. Since accretion rate
$\dot{M}_a\propto T_{z\phi}^{\rm Max}$, increasing vertical net field reduces the
ratio of $\dot{M}_w/\dot{M}_a$. This is accompanied by the outward shift of the
location of the Alfv\'en point with increasing net vertical flux, and for
$\beta_0=10^3$, the Alfv\'en point is beyond our simulation box.
We also note for relatively large net vertical flux of $\beta_0=10^3$, the
wind-driven accretion rate (Equation \ref{eq:accrate}) would become too large
compared with observations. Therefore, a reasonable wind-driven scenario
in PPDs should involve only weak vertical background field with $\beta_0$ of the
order $10^6$ to a few times $10^4$.

Increasing the net vertical field also leads to rapid increase in the midplane
magnetic field (undead zone), which leads to larger $\alpha_{\rm Max}$. Applying
Equation (\ref{eq:accrate}), we find the radial transport remains to play a minor
role to the total accretion rate for all value of $\beta_0$ explored so far, but its
contribution increases with increasing net flux.


Finally, we find that for relatively strong net vertical field with $\beta_0\lesssim10^4$,
it is less likely to obtain a laminar solution with the physical wind geometry: the
strong current layer would escape from our simulation box and one recovers the
undesired solution with odd-$z$ symmetry. This observation may suggest that
maintaining the stability of the strong current layer becomes more difficult for larger
field strength, although the issue can only be fully resolved by performing global
simulations.

\subsection{Effect of FUV Ionization}\label{ssec:fuv}

The FUV ionization is another crucial ingredient of wind launching: the large
ionization fraction in the FUV layer makes the gas and magnetic field be strongly
coupled to each other so that it is essentially in the ideal MHD regime. The
strong coupling between the gas and magnetic field is essential for effectively
loading mass onto open magnetic field lines for magneto-centrifugal acceleration
to take place. Therefore, we expect the rate of mass outflow to strongly depend
on the penetration depth of the FUV ionization. Indeed, we see from Figure
\ref{fig:parameter} that increasing $\Sigma_{\rm FUV}$ by a factor of 10
results in a factor of more than 3 increase of the mass outflow rate. Meanwhile,
increasing $\Sigma_{\rm FUV}$ also leads to a moderate increase of the wind
stress, as seen on the right panel of Figure \ref{fig:parameter}.
Correspondingly, the ratio of $\dot{M}_w/\dot{M}_a$ increases with
$\Sigma_{\rm FUV}$, in parallel with the lowering of the the Alfv\'en point.

\subsection{Effect of Grain Abundance}

In the MRI-driven accretion scenario studied in \citet{Bai11a}, the predicted
accretion rate sensitively depends on the size and abundance of grains. In the
wind-driven accretion scenario, we find that the dependence on grains is much
weaker. For the outflow rate, there is only a $50\%$ difference between the
grain-free chemistry and the case with $0.1\mu$m grains with the abundance of
$\epsilon_{\rm gr}=10^{-3}$. The wind stress, on the other hand, is almost
independent of grain abundance. The weak dependence on the grain abundance
is mainly due to the fact that the wind launching process mainly depends on the
gas conductivity in the surface layer of the disk where grains only play a minor role:
either the FUV ionization dominates (which is independent of
grain abundance), or the long recombination time in the low gas density leads to
large ionization fraction that well exceeds the grain abundance (hence grains
have only very limited effect on the ionization level, see Figure 1 of \citealp{Bai11a}).
The grain abundance does have a strong effect on the radial transport, as we see
from Table \ref{tab:windresults} that $\alpha_{\rm Max}$ increases relatively rapidly
with decreasing grain abundance. Nevertheless, the overall picture is unchanged
since radial transport only plays a minor role in driving disk accretion.

\subsection[]{Effect of Disk Mass}

The surface density of the inner PPDs is highly uncertain and have not been
well constrained observationally (situation is better for the outer disk with
millimeter interferometric observations, e.g., \citealp{Andrews_etal09}). The
MMSN surface density serves as an initial guess and is likely on the lower
end, while much more massive inner disks are also speculated (e.g.,
\citealp{Desch07}). On the other hand, the surface density should decrease
with time as the disks evolve. Therefore, we further consider disk models
that are three time (with label ``M3") and $0.3$ times (with label ``M03")
the mass of the fiducial MMSN disk, while we have also varied the net vertical
flux in code unit so that the physical magnetic fluxes are the same for all runs.
Because we have adopted midplane density $\rho_0=1$ as our natural unit
whereas in reality $\rho_0\propto\Sigma$, hence some conversion is
necessary. In Table \ref{tab:windresults2}, we see that the wind fluxes
$T_{z\phi}^{\rm Max}$ in the two cases vary by a factor of about $10$, and
the fiducial case lies in between. This means that in physical unit, the stress
is more or less independent of the disk surface density. The wind mass loss
rate $\dot{M}_w$ behaves similarly and is roughly independent of $\Sigma$.
Therefore, the accretion and mass loss rate in PPDs is largely determined
by the poloidal magnetic flux distribution through the disk, regardless of disk
mass.

\section[]{Discussions and Conclusion}\label{sec:discussion}

\subsection[]{Summary}

In this paper, we have performed three-dimensional shearing-box simulations
to study the gas dynamics in the inner region of a PPD (at 1 AU). Non-ideal
MHD effects, namely the Ohmic resistivity, Hall effect and ambipolar diffusion
(AD) are crucial in the weakly ionized gas as in PPDs. They affect the coupling
between the gas and magnetic field in different ways and dramatically change
the MHD stability of the system. Particularly, Ohmic resistivity dominates
in the midplane region while AD dominates in the disk surface layer, and the Hall
dominated regime lies in between. For the first time, we have included both
Ohmic resistivity and AD in vertically stratified simulations. The diffusion
coefficients are obtained self-consistently by interpolating from a pre-computed
look-up table (as a function of density and ionization rate at fixed temperature)
based on equilibrium chemistry.

We first showed that the conventional picture of layered accretion fails when
AD is taken into account. Without considering AD (the conventional scenario),
the MRI drives vigorous turbulence in the disk surface (active) layer that easily
accounts for the observed accretion rates in PPDs. However, we find that the
MRI activity is severely suppressed by AD, and the gas dynamics also
sensitively depends on magnetic field geometry. More specifically, 
in the zero net vertical flux field configuration, the disk exhibits extremely weak
turbulent activity, with total stress parameter $\alpha\sim3\times10^{-6}$, mostly
contributed from a thin layer above the FUV ionization front. The resulting
steady state accretion rate is about three orders of magnitude too small
compared with observations. The main reason for such inefficiency is that MRI
in the AD dominated regime disfavors zero net vertical flux field geometry
\citep{BaiStone11}. By contrast, in the presence of net vertical magnetic flux,
the MRI turbulent activity disappears and the flow is completely laminar. In the
mean time, the disk launches a strong wind, which is accelerated to
super-Alfv\'enic velocities within our simulation domain. Instead of MRI driven
accretion in the conventional scenario, angular momentum is carried away by
the magnetocentrifugal wind.

We provide detailed analysis of the laminar wind solution. The wind launching
process is mainly assisted by the vertical gradient of toroidal magnetic field,
starting from about $z\sim\pm2H$ where gas and field lines are marginally
coupled to each other (near the midplane they are essentially decoupled). At
the base of the wind, the azimuthal gas velocity exceeds Keplerian, and the
magnetocentrifugal mechanism takes place and efficiently accelerates the
wind flow. We find that the magnetocentrifugal wind very effectively carries
away angular momentum from the disk. Even with a small net vertical magnetic
flux of $\beta_0=10^5$, the launched wind is already sufficient to account for the
observed accretion rate of about $10^{-8}M_{\bigodot}$ yr$^{-1}$. The rate of the
mass outflow $\dot{M}_w$ is not a well characterized quantity in shearing-box
simulations, which decreases with increasing vertical domain size. We
speculate that realistic mass loss rate is much smaller than that measured in
our simulations based on studying the dependence of $\dot{M}_w$ on box
height; on the other hand, the wind stress, hence the wind-driven accretion
rate increases only slightly with increasing box height, hence our measurement
is more reliable.

The natural symmetry of the wind solution obtained in our shearing-box
simulations  has an odd-$z$ symmetry which is inconsistent with a physical
wind geometry. Meanwhile, we find that the system has an equal probability
to evolve into an exactly same solution but with horizontal velocity and
magnetic field flipped at one side of the box. The flip takes place in
a strong current layer with a thickness of $\sim0.25H$ that is offset from the
disk midplane (due to the large resistivity at midplane). This second solution
has the desired physical geometry of the magnetocentrifugal wind, and in this
solution, the entire accretion flow is carried within the strong current layer, with
radial velocity achieving $\sim0.1-0.3c_s$, while the rest of the disk remains
static. We further show that a solution with enforced even-$z$ symmetry gives
exactly the same magnetocentrifugal wind in the disk upper layer, which
further verifies the robustness of wind-driven accretion scenario.

We have performed a parameter study where we show that the rate of wind
angular momentum transport $T_{z\phi}^{\rm Max}$ increases rapidly with
increasing net vertical flux, with approximately
$T_{z\phi}^{\rm Max}\propto \beta_0^{-0.7}$. Similarly,
$\dot{M}_w\propto\beta_0^{-0.5}$. Moreover, both the wind stress and the
wind mass loss rate depend only on the physical vertical field strength and
very weakly depends on the surface density of the disk. The FUV ionization
which makes the gas in the wind zone behave as ideal MHD, is essential
for wind mass loading. The depth of the FUV ionization, which
depends on the extent of dust attenuation and self-shielding, also affects the
efficiency of wind transport of angular momentum and wind mass loss
relatively strongly. Finally, we find that the grain abundance, which has
been found to play a crucial role in determining the extent of the active layer
in the conventional layered accretion scenario
\citep{IlgnerNelson06,BaiGoodman09}, has almost no influence on the wind
scenario. This is mainly because that the wind launching process takes
place in the upper layer of the disk where the chemistry is less affected by
the grains (since ionization fraction well exceeds grain abundance).

\subsection{Implications}\label{ssec:imp}

The fact that the inner region of PPDs is likely to be laminar with purely wind-driven
accretion processing through a strong current layer offset from the disk midplane
may have profound implications on many aspects of planet formation, particularly
on the following two aspects.

The laminar inner disk is likely to become the mostly favored spot for grain growth
and settling, as well as planetesimal formation and growth. In the conventional picture
of layered accretion, the Ohmic dead zone region is found to be not completely static,
but has random motion  due to the sound waves launched from the active layers
\citep{FlemingStone03,OishiMacLow09,Turner_etal10,OkuzumiHirose11}. The
level of the random motion, albeit much smaller than that in the conventional
active layers, can still be large enough to prevent the solids from settling
completely. The random gas velocities plays an important role on the collision
velocity between dust grains that would impede grain growth \citep{OrmelCuzzi07}.
Moreover, the stochastic gravitational torque from density fluctuations would excite
the random velocities among existing planetesimals to sufficiently large velocities
that their mutual collisions may lead to fragmentation rather than net growth
\citep{Ida_etal08,Yang_etal12}. More recently, \citet{Gressel_etal12} and
\citet{OkuzumiHirose12} found that increasing the vertical net magnetic flux
strongly increases the level of random gas motion as well as the strength of the
density waves in the midplane hence greatly reduces the maximum grain size
achievable from grain growth, and may lead to collisional destruction of
planetesimals. In our laminar wind solution, the disk midplane is essentially
completely static. This provides the best environment for grain growth and settling,
and is ideal for planetesimal formation and survival either via the streaming
instability mechanism \citep{Johansen_etal09,BaiStone10b}, or the gravitational
instability mechanism \citep{Lee_etal10b,Youdin11}, or via the vortices generated by
the baroclinic instability \citep{LesurPap10,LyraKlahr11} or the Rossby wave
instability \citep{Lovelace_etal99,Meheut_etal12a}.

The efficient wind-driven accretion through the inner disk would change
our understanding about the global evolution of PPDs. Most current models on
the long-term evolution of PPDs adopt a phenomenological approach that treats
the disk physics very roughly \citep{Zhu_etal10b,Martin_etal12}. In particular,
under the framework of layered accretion in the inner disk, the presence of the
dead zone often leads to inefficient accretion and pile-up of mass. The pile-up
gradually leads to gravitational instability which eventually drains most of the
inner disk material onto the star, starting a new cycle. In the mean time, zero
net vertical flux global simulations by \citet{Dzyurkevich_etal10} found that the
inner edge of the dead zone is likely to become a local pressure maxima which
serves to trap solids. In reality, when more microphysics in the disk is taken
into account, we see that as disk wind can be much more efficient in driving
accretion at the location of the conventional dead zones, mass pile-up and
building up a local pressure maxima may be avoided. What determines the
surface density profile of the disk is the spatial distribution of the magnetic flux,
rather than the grain abundance, etc. Although our knowledge on the magnetic
flux distribution in PPDs is very limited \citep{Zhao_etal11,Krasnopolsky_etal12},
and it sensitively depends on the internal structure of the disk \citep{GuiletOgilvie13},
attention should be drawn to re-consider the structure and evolution of the
PPDs focusing more on the transport of large-scale magnetic fields using global
simulations.

\subsection{Open Issues}\label{ssec:open}

We have neglected the Hall effect in our calculation, which also plays an
important role on the gas dynamics in PPDs. In the bottom left panel of Figure
\ref{fig:FidOA-prof}, we also show the profile of the Hall Elsasser number for
our laminar wind solution. We see that the Hall effect at least dominates
between $z=2H$ and $3H$ (within $z=\pm2H$, a floor is added to Ohmic
resistivity and AD, but not to the Hall term since it is not included in the
computation). At this location, we have $Ha\sim0.1$, $\beta\sim10$, and the
azimuthal magnetic field dominates the vertical field. We note that for the Hall
MRI, the stability depends on the sign of vertical magnetic field, while for Ohmic
resistivity and AD, stability is independent of the sign of $B_z$. We expect the
magnetic configuration in our laminar wind solution to be stable to the Hall MRI
when $B_z$ is negative, in which case we expect the magnetic configuration to
be adjusted to account for the Hall effect. For positive $B_z$, our laminar
solution may become unstable to the Hall MRI according to the calculations
by \citet{PandeyWardle12}, but the unstable wavelength may well exceed $H$
due to the small value of plasma $\beta$. Moreover, as vertical stratification is
not included in their calculation, it is uncertain whether the disk would eventually
relax to another laminar configuration or lead to sustained MRI turbulence,
especially given the experience in this work that an initially MRI unstable disk
evolves into a laminar configuration in non-linear simulations.

We have shown the the FUV ionization plays an important role in the wind
launching process. In the mean time, the FUV photons are also important
dirver of photoevaporation \citep{GortiHollenbach09,Owen_etal12}, which
is not considered in this work. Photoevaporation is considered as the main
mechanism for disk dispersal and largely determines the lifetime of PPDs
\citep{Alexander_etal06b}. We expect photoevaporation and magnetocentrifugal
mechanisms to boost each other and enhance the mass loss rate compared with
pure photoevaporation models, yet more realistic thermodynamics with proper
treatment of heating and cooling need to be included, and eventually to
make predictions to compare with observations \citep{Pascucci_etal09,Sacco_etal12}.

The magneto-centrifugal wind is intrinsically a global phenomenon. As we
have discussed before, uncertainties remains in our local shearing-box
simulations regarding problems with the stability of the strong current layer,
the mass loss rate, and the rate of wind-driven angular momentum transport.
Moreover, we have only studied the gas dynamics for a MMSN disk at 1 AU,
it is yet to conduct the same study at other disk locations to study how the
wind properties depend on the radius. This will be presented in a companion
paper, where we will show that the laminar wind scenario is likely to extend to
about 10 AU, while the MRI (in the presence of net vertical magnetic flux)
is likely to be the dominant mechanism for angular momentum transport at the
outer disk \citep{Simon_etal13}. Meanwhile,
another interesting problem arises on how the transition occurs from a pure
laminar wind-driven accretion region to a turbulent MRI-driven accretion
region. Global simulations are essential to resolve the problems and
uncertainties mentioned above, and are the ultimate way towards fully
understanding the accretion processes in PPDs.

\acknowledgments

We thank Jeremy Goodman, Arieh Konigl and Neal Turner for useful
discussions. XNB acknowledges support for program number HST-HF-51301.01-A
provided by NASA through a Hubble Fellowship grant from the Space Telescope
Science Institute, which is operated by the Association of Universities for Research
in Astronomy, Incorporated, under NASA contract NAS5-26555.
JMS acknowledges support from the National Science Foundation through grant
AST-0908269.
This work used computational facilities provided by PICSciE at Princeton University.
Part of the computation is performed on Kraken and Nautilus at the National Institute
for Computational Sciences through XSEDE grant TG-AST090106.



\begin{thebibliography}{125}
\expandafter\ifx\csname natexlab\endcsname\relax\def\natexlab#1{#1}\fi

\bibitem[{{{\'A}d{\'a}mkovics} {et~al.}(2011){{\'A}d{\'a}mkovics}, {Glassgold},
  \& {Meijerink}}]{Adamkovics_etal11}
{{\'A}d{\'a}mkovics}, M., {Glassgold}, A.~E., \& {Meijerink}, R. 2011, \apj,
  736, 143

\bibitem[{{Alexander} {et~al.}(2006){Alexander}, {Clarke}, \&
  {Pringle}}]{Alexander_etal06b}
{Alexander}, R.~D., {Clarke}, C.~J., \& {Pringle}, J.~E. 2006, \mnras, 369, 229

\bibitem[{{Anderson} {et~al.}(2005){Anderson}, {Li}, {Krasnopolsky}, \&
  {Blandford}}]{Anderson_etal05}
{Anderson}, J.~M., {Li}, Z.-Y., {Krasnopolsky}, R., \& {Blandford}, R.~D. 2005,
  \apj, 630, 945

\bibitem[{{Andrews} {et~al.}(2009){Andrews}, {Wilner}, {Hughes}, {Qi}, \&
  {Dullemond}}]{Andrews_etal09}
{Andrews}, S.~M., {Wilner}, D.~J., {Hughes}, A.~M., {Qi}, C., \& {Dullemond},
  C.~P. 2009, \apj, 700, 1502

\bibitem[{{Bai}(2011{\natexlab{a}})}]{Bai11a}
{Bai}, X.-N. 2011{\natexlab{a}}, \apj, 739, 50

\bibitem[{{Bai}(2011{\natexlab{b}})}]{Bai11b}
---. 2011{\natexlab{b}}, \apj, 739, 51

\bibitem[{{Bai} \& {Goodman}(2009)}]{BaiGoodman09}
{Bai}, X.-N. \& {Goodman}, J. 2009, \apj, 701, 737

\bibitem[{{Bai} \& {Stone}(2010{\natexlab{a}})}]{BaiStone10b}
{Bai}, X.-N. \& {Stone}, J.~M. 2010{\natexlab{a}}, \apj, 722, 1437

\bibitem[{{Bai} \& {Stone}(2010{\natexlab{b}})}]{BaiStone10c}
---. 2010{\natexlab{b}}, \apjl, 722, L220

\bibitem[{{Bai} \& {Stone}(2011)}]{BaiStone11}
---. 2011, \apj, 736, 144

\bibitem[{{Bai} \& {Stone}(2013)}]{BaiStone12a}
---. 2013, \apj, 767, 30

\bibitem[{{Balbus} \& {Hawley}(1998)}]{BH98}
{Balbus}, S.~A. \& {Hawley}, J.~F. 1998, Reviews of Modern Physics, 70, 1

\bibitem[{{Balbus} \& {Terquem}(2001)}]{BalbusTerquem01}
{Balbus}, S.~A. \& {Terquem}, C. 2001, \apj, 552, 235

\bibitem[{{Bans} \& {K{\"o}nigl}(2012)}]{BansKonigl12}
{Bans}, A. \& {K{\"o}nigl}, A. 2012, \apj, 758, 100

\bibitem[{{Baruteau} {et~al.}(2011){Baruteau}, {Fromang}, {Nelson}, \&
  {Masset}}]{Baruteau_etal11}
{Baruteau}, C., {Fromang}, S., {Nelson}, R.~P., \& {Masset}, F. 2011, \aap,
  533, A84

\bibitem[{{Birnstiel} {et~al.}(2010){Birnstiel}, {Dullemond}, \&
  {Brauer}}]{Birnstiel_etal10}
{Birnstiel}, T., {Dullemond}, C.~P., \& {Brauer}, F. 2010, \aap, 513, A79+

\bibitem[{{Blaes} \& {Balbus}(1994)}]{BlaesBalbus94}
{Blaes}, O.~M. \& {Balbus}, S.~A. 1994, \apj, 421, 163

\bibitem[{{Blandford} \& {Payne}(1982)}]{BlandfordPayne82}
{Blandford}, R.~D. \& {Payne}, D.~G. 1982, \mnras, 199, 883

\bibitem[{{Cabrit} {et~al.}(1990){Cabrit}, {Edwards}, {Strom}, \&
  {Strom}}]{Cabrit_etal90}
{Cabrit}, S., {Edwards}, S., {Strom}, S.~E., \& {Strom}, K.~M. 1990, \apj, 354,
  687

\bibitem[{{Casse} \& {Keppens}(2002)}]{CasseKeppens02}
{Casse}, F. \& {Keppens}, R. 2002, \apj, 581, 988

\bibitem[{{Casse} \& {Keppens}(2004)}]{CasseKeppens04}
---. 2004, \apj, 601, 90

\bibitem[{{Combet} \& {Ferreira}(2008)}]{CombetFerreira08}
{Combet}, C. \& {Ferreira}, J. 2008, \aap, 479, 481

\bibitem[{{Davis} {et~al.}(2010){Davis}, {Stone}, \& {Pessah}}]{Davis_etal10}
{Davis}, S.~W., {Stone}, J.~M., \& {Pessah}, M.~E. 2010, \apj, 713, 52

\bibitem[{{Desch}(2004)}]{Desch04}
{Desch}, S.~J. 2004, \apj, 608, 509

\bibitem[{{Desch}(2007)}]{Desch07}
---. 2007, \apj, 671, 878

\bibitem[{{Dzyurkevich} {et~al.}(2010){Dzyurkevich}, {Flock}, {Turner},
  {Klahr}, \& {Henning}}]{Dzyurkevich_etal10}
{Dzyurkevich}, N., {Flock}, M., {Turner}, N.~J., {Klahr}, H., \& {Henning}, T.
  2010, \aap, 515, A70+
  
\bibitem[Dzyurkevich et al.(2013)]{Dzyurkevich_etal13} Dzyurkevich, N., 
Turner, N.~J., Henning, T., \& Kley, W.\ 2013, \apj, 765, 114 

\bibitem[{{Ferreira} \& {Pelletier}(1995)}]{FerreiraPelletier95}
{Ferreira}, J. \& {Pelletier}, G. 1995, \aap, 295, 807

\bibitem[{{Flaig} {et~al.}(2012){Flaig}, {Ruoff}, {Kley}, \&
  {Kissmann}}]{Flaig_etal12}
{Flaig}, M., {Ruoff}, P., {Kley}, W., \& {Kissmann}, R. 2012, \mnras, 420, 2419

\bibitem[{{Fleming} \& {Stone}(2003)}]{FlemingStone03}
{Fleming}, T. \& {Stone}, J.~M. 2003, \apj, 585, 908

\bibitem[{{Fleming} {et~al.}(2000){Fleming}, {Stone}, \&
  {Hawley}}]{Fleming_etal00}
{Fleming}, T.~P., {Stone}, J.~M., \& {Hawley}, J.~F. 2000, \apj, 530, 464

\bibitem[{{Flock} {et~al.}(2012){Flock}, {Henning}, \& {Klahr}}]{Flock_etal12}
{Flock}, M., {Henning}, T., \& {Klahr}, H. 2012, \apj, 761, 95

\bibitem[{{Fromang} {et~al.}(2012){Fromang}, {Latter}, {Lesur}, \&
  {Ogilvie}}]{Fromang_etal12}
{Fromang}, S., {Latter}, H.~N., {Lesur}, G., \& {Ogilvie}, G.~I. 2013, \aap,
  552, A71

\bibitem[{{Gammie}(1996)}]{Gammie96}
{Gammie}, C.~F. 1996, \apj, 457, 355

\bibitem[{{Garaud}(2007)}]{Garaud07}
{Garaud}, P. 2007, \apj, 671, 2091

\bibitem[{{Goldreich} \& {Lynden-Bell}(1965)}]{GoldreichLyndenBell65}
{Goldreich}, P. \& {Lynden-Bell}, D. 1965, \mnras, 130, 125

\bibitem[{{Gorti} \& {Hollenbach}(2009)}]{GortiHollenbach09}
{Gorti}, U. \& {Hollenbach}, D. 2009, \apj, 690, 1539

\bibitem[{{Gressel} {et~al.}(2012){Gressel}, {Nelson}, \&
  {Turner}}]{Gressel_etal12}
{Gressel}, O., {Nelson}, R.~P., \& {Turner}, N.~J. 2012, \mnras, 422, 1140

\bibitem[{{Guilet} \& {Ogilvie}(2012)}]{GuiletOgilvie13}
{Guilet}, J. \& {Ogilvie}, G.~I. 2013, \mnras, 430, 822

\bibitem[{{Hartigan} {et~al.}(1995){Hartigan}, {Edwards}, \&
  {Ghandour}}]{Hartigan_etal95}
{Hartigan}, P., {Edwards}, S., \& {Ghandour}, L. 1995, \apj, 452, 736

\bibitem[{{Hartmann} {et~al.}(1998){Hartmann}, {Calvet}, {Gullbring}, \&
  {D'Alessio}}]{Hartmann_etal98}
{Hartmann}, L., {Calvet}, N., {Gullbring}, E., \& {D'Alessio}, P. 1998, \apj,
  495, 385

\bibitem[{{Hawley} {et~al.}(1995){Hawley}, {Gammie}, \& {Balbus}}]{HGB95}
{Hawley}, J.~F., {Gammie}, C.~F., \& {Balbus}, S.~A. 1995, \apj, 440, 742

\bibitem[{{Hawley} \& {Stone}(1998)}]{HawleyStone98}
{Hawley}, J.~F. \& {Stone}, J.~M. 1998, \apj, 501, 758

\bibitem[{{Hayashi}(1981)}]{Hayashi81}
{Hayashi}, C. 1981, Progress of Theoretical Physics Supplement, 70, 35

\bibitem[{{Heinzeller} {et~al.}(2011){Heinzeller}, {Nomura}, {Walsh}, \&
  {Millar}}]{Heizeller_etal11}
{Heinzeller}, D., {Nomura}, H., {Walsh}, C., \& {Millar}, T.~J. 2011, \apj,
  731, 115

\bibitem[{{Hirose} \& {Turner}(2011)}]{HiroseTurner11}
{Hirose}, S. \& {Turner}, N.~J. 2011, \apjl, 732, L30

\bibitem[{{Hughes} \& {Armitage}(2010)}]{HughesArmitage10}
{Hughes}, A.~L.~H. \& {Armitage}, P.~J. 2010, \apj, 719, 1633

\bibitem[{{Ida} {et~al.}(2008){Ida}, {Guillot}, \& {Morbidelli}}]{Ida_etal08}
{Ida}, S., {Guillot}, T., \& {Morbidelli}, A. 2008, \apj, 686, 1292

\bibitem[{{Igea} \& {Glassgold}(1999)}]{IG99}
{Igea}, J. \& {Glassgold}, A.~E. 1999, \apj, 518, 848

\bibitem[{{Ilgner} \& {Nelson}(2006)}]{IlgnerNelson06}
{Ilgner}, M. \& {Nelson}, R.~P. 2006, \aap, 445, 205

\bibitem[{{Ilgner} \& {Nelson}(2008)}]{IlgnerNelson08}
---. 2008, \aap, 483, 815

\bibitem[{{Johansen} {et~al.}(2009){Johansen}, {Youdin}, \& {Mac
  Low}}]{Johansen_etal09}
{Johansen}, A., {Youdin}, A., \& {Mac Low}, M. 2009, \apjl, 704, L75

\bibitem[{{Kato} {et~al.}(2002){Kato}, {Kudoh}, \& {Shibata}}]{Kato_etal02}
{Kato}, S.~X., {Kudoh}, T., \& {Shibata}, K. 2002, \apj, 565, 1035

\bibitem[{{K{\"o}nigl} {et~al.}(2010){K{\"o}nigl}, {Salmeron}, \&
  {Wardle}}]{Konigl_etal10}
{K{\"o}nigl}, A., {Salmeron}, R., \& {Wardle}, M. 2010, \mnras, 401, 479

\bibitem[{{Krasnopolsky} {et~al.}(1999){Krasnopolsky}, {Li}, \&
  {Blandford}}]{Krasnopolsky_etal99}
{Krasnopolsky}, R., {Li}, Z.-Y., \& {Blandford}, R. 1999, \apj, 526, 631

\bibitem[{{Krasnopolsky} {et~al.}(2003){Krasnopolsky}, {Li}, \&
  {Blandford}}]{Krasnopolsky_etal03}
{Krasnopolsky}, R., {Li}, Z.-Y., \& {Blandford}, R.~D. 2003, \apj, 595, 631

\bibitem[{{Krasnopolsky} {et~al.}(2012){Krasnopolsky}, {Li}, {Shang}, \&
  {Zhao}}]{Krasnopolsky_etal12}
{Krasnopolsky}, R., {Li}, Z.-Y., {Shang}, H., \& {Zhao}, B. 2012, \apj, 757, 77

\bibitem[{{Kunz} \& {Balbus}(2004)}]{KunzBalbus04}
{Kunz}, M.~W. \& {Balbus}, S.~A. 2004, \mnras, 348, 355

\bibitem[{{Lee} {et~al.}(2010){Lee}, {Chiang}, {Asay-Davis}, \&
  {Barranco}}]{Lee_etal10b}
{Lee}, A.~T., {Chiang}, E., {Asay-Davis}, X., \& {Barranco}, J. 2010, \apj,
  725, 1938

\bibitem[{{Lesur} {et~al.}(2012){Lesur}, {Ferreira}, \&
  {Ogilvie}}]{Lesur_etal12}
{Lesur}, G., {Ferreira}, J., \& {Ogilvie}, G.~I. 2013, \aap, 550, A61

\bibitem[{{Lesur} \& {Papaloizou}(2010)}]{LesurPap10}
{Lesur}, G. \& {Papaloizou}, J.~C.~B. 2010, \aap, 513, A60

\bibitem[{{Li}(1995)}]{Li95}
{Li}, Z.-Y. 1995, \apj, 444, 848

\bibitem[{{Li}(1996)}]{Li96}
---. 1996, \apj, 465, 855

\bibitem[{{Lovelace} {et~al.}(1999){Lovelace}, {Li}, {Colgate}, \&
  {Nelson}}]{Lovelace_etal99}
{Lovelace}, R.~V.~E., {Li}, H., {Colgate}, S.~A., \& {Nelson}, A.~F. 1999,
  \apj, 513, 805

\bibitem[{{Lyra} \& {Klahr}(2011)}]{LyraKlahr11}
{Lyra}, W. \& {Klahr}, H. 2011, \aap, 527, A138

\bibitem[{{Martin} {et~al.}(2012){Martin}, {Lubow}, {Livio}, \&
  {Pringle}}]{Martin_etal12}
{Martin}, R.~G., {Lubow}, S.~H., {Livio}, M., \& {Pringle}, J.~E. 2012, \mnras,
  420, 3139

\bibitem[{{Meheut} {et~al.}(2012){Meheut}, {Yu}, \& {Lai}}]{Meheut_etal12a}
{Meheut}, H., {Yu}, C., \& {Lai}, D. 2012, \mnras, 422, 2399

\bibitem[{{Miller} \& {Stone}(2000)}]{MillerStone00}
{Miller}, K.~A. \& {Stone}, J.~M. 2000, \apj, 534, 398

\bibitem[Mohanty et al.(2013)]{Mohanty_etal13} Mohanty, S., Ercolano, 
B., \& Turner, N.~J.\ 2013, \apj, 764, 65

\bibitem[{{Ogilvie}(2012)}]{Ogilvie12}
{Ogilvie}, G.~I. 2012, \mnras, 423, 1318

\bibitem[{{Ogilvie} \& {Livio}(2001)}]{OgilvieLivio01}
{Ogilvie}, G.~I. \& {Livio}, M. 2001, \apj, 553, 158

\bibitem[{{Oishi} \& {Mac Low}(2009)}]{OishiMacLow09}
{Oishi}, J.~S. \& {Mac Low}, M. 2009, \apj, 704, 1239

\bibitem[{{Okuzumi} \& {Hirose}(2011)}]{OkuzumiHirose11}
{Okuzumi}, S. \& {Hirose}, S. 2011, \apj, 742, 65

\bibitem[{{Okuzumi} \& {Hirose}(2012)}]{OkuzumiHirose12}
---. 2012, \apjl, 753, L8

\bibitem[{{Ormel} \& {Cuzzi}(2007)}]{OrmelCuzzi07}
{Ormel}, C.~W. \& {Cuzzi}, J.~N. 2007, \aap, 466, 413

\bibitem[{{Ouyed} \& {Pudritz}(1997)}]{OuyedPudritz97a}
{Ouyed}, R. \& {Pudritz}, R.~E. 1997, \apj, 482, 712

\bibitem[{{Ouyed} {et~al.}(1997){Ouyed}, {Pudritz}, \& {Stone}}]{Ouyed_etal97}
{Ouyed}, R., {Pudritz}, R.~E., \& {Stone}, J.~M. 1997, \nat, 385, 409

\bibitem[{{Owen} {et~al.}(2012){Owen}, {Clarke}, \& {Ercolano}}]{Owen_etal12}
{Owen}, J.~E., {Clarke}, C.~J., \& {Ercolano}, B. 2012, \mnras, 422, 1880

\bibitem[{{Paardekooper} {et~al.}(2010){Paardekooper}, {Baruteau}, {Crida}, \&
  {Kley}}]{Paardekooper_etal10}
{Paardekooper}, S.-J., {Baruteau}, C., {Crida}, A., \& {Kley}, W. 2010, \mnras,
  401, 1950

\bibitem[{{Pandey} \& {Wardle}(2012)}]{PandeyWardle12}
{Pandey}, B.~P. \& {Wardle}, M. 2012, \mnras, 3001

\bibitem[{{Pascucci} \& {Sterzik}(2009)}]{Pascucci_etal09}
{Pascucci}, I. \& {Sterzik}, M. 2009, \apj, 702, 724

\bibitem[{{Pelletier} \& {Pudritz}(1992)}]{PelletierPudritz92}
{Pelletier}, G. \& {Pudritz}, R.~E. 1992, \apj, 394, 117

\bibitem[{{Perez-Becker} \&
  {Chiang}(2011{\natexlab{a}})}]{PerezBeckerChiang11a}
{Perez-Becker}, D. \& {Chiang}, E. 2011{\natexlab{a}}, \apj, 727, 2

\bibitem[{{Perez-Becker} \&
  {Chiang}(2011{\natexlab{b}})}]{PerezBeckerChiang11b}
---. 2011{\natexlab{b}}, \apj, 735, 8

\bibitem[{{Pudritz} \& {Norman}(1983)}]{PudritzNorman83}
{Pudritz}, R.~E. \& {Norman}, C.~A. 1983, \apj, 274, 677

\bibitem[{{Sacco} {et~al.}(2012){Sacco}, {Flaccomio}, {Pascucci}, {Lahuis},
  {Ercolano}, {Kastner}, {Micela}, {Stelzer}, \& {Sterzik}}]{Sacco_etal12}
{Sacco}, G.~G., {Flaccomio}, E., {Pascucci}, I., {Lahuis}, F., {Ercolano}, B.,
  {Kastner}, J.~H., {Micela}, G., {Stelzer}, B., \& {Sterzik}, M. 2012, \apj,
  747, 142

\bibitem[{{Salmeron} {et~al.}(2007){Salmeron}, {K{\"o}nigl}, \&
  {Wardle}}]{Salmeron_etal07}
{Salmeron}, R., {K{\"o}nigl}, A., \& {Wardle}, M. 2007, \mnras, 375, 177

\bibitem[{{Salmeron} {et~al.}(2011){Salmeron}, {K{\"o}nigl}, \&
  {Wardle}}]{Salmeron_etal11}
---. 2011, \mnras, 412, 1162

\bibitem[{{Salmeron} \& {Wardle}(2005)}]{SalmeronWardle05}
{Salmeron}, R. \& {Wardle}, M. 2005, \mnras, 361, 45

\bibitem[{{Salmeron} \& {Wardle}(2008)}]{SalmeronWardle08}
---. 2008, \mnras, 388, 1223

\bibitem[{{Sano} \& {Stone}(2002{\natexlab{a}})}]{SanoStone02a}
{Sano}, T. \& {Stone}, J.~M. 2002{\natexlab{a}}, \apj, 570, 314

\bibitem[{{Sano} \& {Stone}(2002{\natexlab{b}})}]{SanoStone02b}
---. 2002{\natexlab{b}}, \apj, 577, 534

\bibitem[{{Semenov} {et~al.}(2006){Semenov}, {Wiebe}, \&
  {Henning}}]{Semenov_etal06}
{Semenov}, D., {Wiebe}, D., \& {Henning}, T. 2006, \apjl, 647, L57

\bibitem[{{Shakura} \& {Sunyaev}(1973)}]{ShakuraSunyaev73}
{Shakura}, N.~I. \& {Sunyaev}, R.~A. 1973, \aap, 24, 337

\bibitem[{{Shi} {et~al.}(2010){Shi}, {Krolik}, \& {Hirose}}]{Shi_etal10}
{Shi}, J., {Krolik}, J.~H., \& {Hirose}, S. 2010, \apj, 708, 1716

\bibitem[{{Shu} {et~al.}(1994){Shu}, {Najita}, {Ostriker}, {Wilkin}, {Ruden},
  \& {Lizano}}]{Shu_etal94}
{Shu}, F., {Najita}, J., {Ostriker}, E., {Wilkin}, F., {Ruden}, S., \&
  {Lizano}, S. 1994, \apj, 429, 781

\bibitem[{{Shu} {et~al.}(2008){Shu}, {Lizano}, {Galli}, {Cai}, \&
  {Mohanty}}]{Shu_etal08}
{Shu}, F.~H., {Lizano}, S., {Galli}, D., {Cai}, M.~J., \& {Mohanty}, S. 2008,
  \apjl, 682, L121

\bibitem[{{Sicilia-Aguilar} {et~al.}(2006){Sicilia-Aguilar}, {Hartmann},
  {Calvet}, {Megeath}, {Muzerolle}, {Allen}, {D'Alessio}, {Mer{\'{\i}}n},
  {Stauffer}, {Young}, \& {Lada}}]{Sicilia_etal06}
{Sicilia-Aguilar}, A., {Hartmann}, L., {Calvet}, N., {Megeath}, S.~T.,
  {Muzerolle}, J., {Allen}, L., {D'Alessio}, P., {Mer{\'{\i}}n}, B.,
  {Stauffer}, J., {Young}, E., \& {Lada}, C. 2006, \apj, 638, 897

\bibitem[{{Simon} {et~al.}(2013){Simon}, {Bai}, {Stone}, {Armitage}, \&
  {Beckwith}}]{Simon_etal13}
{Simon}, J.~B., {Bai}, X., {Stone}, J.~M., {Armitage}, P.~J., \& {Beckwith}, K.
  2013, \apj, 764, 66

\bibitem[{{Simon} {et~al.}(2012){Simon}, {Beckwith}, \&
  {Armitage}}]{Simon_etal12a}
{Simon}, J.~B., {Beckwith}, K., \& {Armitage}, P.~J. 2012, \mnras, 422, 2685

\bibitem[{{Sorathia} {et~al.}(2012){Sorathia}, {Reynolds}, {Stone}, \&
  {Beckwith}}]{Sorathia_etal12}
{Sorathia}, K.~A., {Reynolds}, C.~S., {Stone}, J.~M., \& {Beckwith}, K. 2012,
  \apj, 749, 189

\bibitem[{{Spruit}(1996)}]{Spruit96}
{Spruit}, H.~C. 1996, in NATO ASIC Proc. 477: Evolutionary Processes in Binary
  Stars, ed. R.~A.~M.~J. {Wijers}, M.~B. {Davies}, \& C.~A. {Tout}, 249--286

\bibitem[{{Stone} \& {Gardiner}(2010)}]{StoneGardiner10}
{Stone}, J.~M. \& {Gardiner}, T.~A. 2010, \apjs, 189, 142

\bibitem[{{Stone} {et~al.}(2008){Stone}, {Gardiner}, {Teuben}, {Hawley}, \&
  {Simon}}]{Stone_etal08}
{Stone}, J.~M., {Gardiner}, T.~A., {Teuben}, P., {Hawley}, J.~F., \& {Simon},
  J.~B. 2008, \apjs, 178, 137

\bibitem[{{Stone} {et~al.}(1996){Stone}, {Hawley}, {Gammie}, \&
  {Balbus}}]{SHGB96}
{Stone}, J.~M., {Hawley}, J.~F., {Gammie}, C.~F., \& {Balbus}, S.~A. 1996,
  \apj, 463, 656

\bibitem[{{Suzuki} \& {Inutsuka}(2009)}]{SuzukiInutsuka09}
{Suzuki}, T.~K. \& {Inutsuka}, S.-i. 2009, \apjl, 691, L49

\bibitem[{{Suzuki} {et~al.}(2010){Suzuki}, {Muto}, \&
  {Inutsuka}}]{Suzuki_etal10}
{Suzuki}, T.~K., {Muto}, T., \& {Inutsuka}, S.-i. 2010, \apj, 718, 1289

\bibitem[{{Teitler}(2011)}]{Teitler11}
{Teitler}, S. 2011, \apj, 733, 57

\bibitem[{{Turner} {et~al.}(2010){Turner}, {Carballido}, \&
  {Sano}}]{Turner_etal10}
{Turner}, N.~J., {Carballido}, A., \& {Sano}, T. 2010, \apj, 708, 188

\bibitem[{{Turner} \& {Sano}(2008)}]{TurnerSano08}
{Turner}, N.~J. \& {Sano}, T. 2008, \apjl, 679, L131

\bibitem[{{Turner} {et~al.}(2007){Turner}, {Sano}, \&
  {Dziourkevitch}}]{Turner_etal07}
{Turner}, N.~J., {Sano}, T., \& {Dziourkevitch}, N. 2007, \apj, 659, 729

\bibitem[{{Umebayashi} \& {Nakano}(1988)}]{UmebayashiNakano88}
{Umebayashi}, T. \& {Nakano}, T. 1988, Progress of Theoretical Physics
  Supplement, 96, 151

\bibitem[{{Walsh} {et~al.}(2012){Walsh}, {Nomura}, {Millar}, \&
  {Aikawa}}]{Walsh_etal12}
{Walsh}, C., {Nomura}, H., {Millar}, T.~J., \& {Aikawa}, Y. 2012, \apj, 747,
  114

\bibitem[{{Wardle}(1997)}]{Wardle97}
{Wardle}, M. 1997, in Astronomical Society of the Pacific Conference Series,
  Vol. 121, IAU Colloq. 163: Accretion Phenomena and Related Outflows, ed.
  D.~T. {Wickramasinghe}, G.~V. {Bicknell}, \& L.~{Ferrario}, 561

\bibitem[{{Wardle}(1999)}]{Wardle99}
{Wardle}, M. 1999, \mnras, 307, 849

\bibitem[{{Wardle}(2007)}]{Wardle07}
---. 2007, \apss, 311, 35

\bibitem[{{Wardle} \& {Koenigl}(1993)}]{WardleKoenigl93}
{Wardle}, M. \& {Koenigl}, A. 1993, \apj, 410, 218

\bibitem[{{Wardle} \& {Salmeron}(2012)}]{WardleSalmeron12}
{Wardle}, M. \& {Salmeron}, R. 2012, \mnras, 422, 2737

\bibitem[{{Weidenschilling}(1977)}]{Weidenschilling77}
{Weidenschilling}, S.~J. 1977, \mnras, 180, 57

\bibitem[{{Williams} \& {Cieza}(2011)}]{WilliamsCieza11}
{Williams}, J.~P. \& {Cieza}, L.~A. 2011, \araa, 49, 67

\bibitem[{{Woodall} {et~al.}(2007){Woodall}, {Ag{\'u}ndez}, {Markwick-Kemper},
  \& {Millar}}]{Woodall_etal07}
{Woodall}, J., {Ag{\'u}ndez}, M., {Markwick-Kemper}, A.~J., \& {Millar}, T.~J.
  2007, \aap, 466, 1197

\bibitem[{{Yang} {et~al.}(2012){Yang}, {Mac Low}, \& {Menou}}]{Yang_etal12}
{Yang}, C.-C., {Mac Low}, M.-M., \& {Menou}, K. 2012, \apj, 748, 79

\bibitem[{{Youdin}(2011)}]{Youdin11}
{Youdin}, A.~N. 2011, \apj, 731, 99

\bibitem[{{Zanni} {et~al.}(2007){Zanni}, {Ferrari}, {Rosner}, {Bodo}, \&
  {Massaglia}}]{Zanni_etal07}
{Zanni}, C., {Ferrari}, A., {Rosner}, R., {Bodo}, G., \& {Massaglia}, S. 2007,
  \aap, 469, 811

\bibitem[{{Zhao} {et~al.}(2011){Zhao}, {Li}, {Nakamura}, {Krasnopolsky}, \&
  {Shang}}]{Zhao_etal11}
{Zhao}, B., {Li}, Z.-Y., {Nakamura}, F., {Krasnopolsky}, R., \& {Shang}, H.
  2011, \apj, 742, 10

\bibitem[{{Zhu} {et~al.}(2010){Zhu}, {Hartmann}, \& {Gammie}}]{Zhu_etal10b}
{Zhu}, Z., {Hartmann}, L., \& {Gammie}, C. 2010, \apj, 713, 1143

\bibitem[{{Zsom} {et~al.}(2010){Zsom}, {Ormel}, {G{\"u}ttler}, {Blum}, \&
  {Dullemond}}]{Zsom_etal10}
{Zsom}, A., {Ormel}, C.~W., {G{\"u}ttler}, C., {Blum}, J., \& {Dullemond},
  C.~P. 2010, \aap, 513, A57+

\end{thebibliography}


\label{lastpage}
\end{document}